\documentstyle[aps,multicol,epsfig]{revtex}  
\begin{document} 
\draft 
\title{Novel Symmetry Classes in
  Mesoscopic Normalconducting-Superconducting Hybrid Structures} 
\author{Alexander Altland  and Martin R. Zirnbauer}
\address{Institut f\"{u}r Theoretische Physik, Universit\"{a}t 
zu K\"{o}ln, Z\"{u}lpicherstr. 77, 50937 K\"{o}ln, Germany} 
\date{Mar 4, 1996} \maketitle

\begin{abstract} Normal-conducting mesoscopic systems in contact with
  a superconductor are classified by the symmetry operations of time
  reversal and rotation of the electron's spin.  Four symmetry classes
  are identified, which correspond to Cartan's symmetric spaces of
  type $C$, $C$I, $D$, and $D$III.  A detailed study is made of the
  systems where the phase shift due to Andreev reflection averages to 
  zero along a typical semiclassical single-electron trajectory.
  Such systems are particularly interesting because they do not have
  a genuine excitation gap but support quasiparticle states close to
  the chemical potential.  Disorder or dynamically generated
  chaos mixes the states and produces novel forms of universal level
  statistics.  For two of the four universality classes, the $n$-level
  correlation functions are calculated by the mapping on a free 1D
  Fermi gas with a boundary.  The remaining two classes are related to
  the Laguerre orthogonal and symplectic random-matrix ensembles.  For
  a quantum dot with an NS-geometry, the weak localization correction
  to the conductance is calculated as a function of sticking
  probability and two perturbations breaking time-reversal
  symmetry and spin-rotation invariance.  The universal conductance
  fluctuations are computed from a maximum-entropy $S$-matrix
  ensemble.  They are larger by a factor of two than what is naively
  expected from the analogy with normal-conducting systems.  This
  enhancement is explained by the doubling of the number of slow
  modes: owing to the coupling of particles and holes by the proximity
  to the superconductor, every cooperon and diffuson mode in the
  advanced-retarded channel entails a corresponding mode in the
  advanced-advanced (or retarded-retarded) channel.  \end{abstract}

\begin{multicols}{2}

\section{Introduction}

Following the early work of Wigner\cite{wigner}, Dyson in his classic
1962 paper\cite{dyson} classified complex many-body systems such as
atomic nuclei according to their fundamental symmetries.  Arguing on
mathematical grounds, he proposed the existence of three symmetry
classes, which are distinguished by their behavior under reversal of
the time direction and by their spin.  The statistical properties of
these classes are described by three random-matrix models, called the
Gaussian Orthogonal, Unitary and Symplectic Ensembles (GOE, GUE, GSE).
Dyson's classification scheme has since proved very far reaching.
Although atomic nuclei display only GOE statistics, physical
realizations of the other two classes were later found in chaotic and
disordered single-electron systems subject to a magnetic field (GUE)
or to spin-orbit scattering (GSE).  

By standard arguments, Wigner-Dyson statistics applies to the {\it 
ergodic} limit, i.e. to times long enough for the degrees of freedom 
to equilibrate and fill the available phase space uniformly.  More 
specifically, in the context of disordered mesoscopic systems the 
ergodic limit is reached for times larger than the diffusion time 
$L^2 / D$, where $D$ is the diffusion constant and $L$ the linear 
extension of the system.  By the uncertainty relation, the ergodic
limit corresponds to the energy range below the Thouless energy 
$\hbar D/L^2$. 

One may ask whether the level statistics of disordered or chaotic 
single-particle systems in the ergodic limit must always be 
Wigner-Dyson or whether different statistics is possible.  The answer 
is that Wigner-Dyson statistics is generic and universal as long as 
the statistics is required to be stationary under shifts of the 
energy.  (This can be understood from the mapping on a nonlinear
$\sigma$ model\cite{efetov}.)  However, if the stationarity
condition is relaxed and additional symmetries are imposed, new 
universality classes may arise.  This happens, for instance, when
a massless Dirac particle is placed in a random gauge field.  
Because the Dirac operator $D$ anticommutes with $\gamma_5$ in the 
chiral (or massless) limit, its matrix is block off-diagonal in the 
eigenbasis of $\gamma_5$.  As a result, the eigenvalues of $D$ are 
either zero or come in pairs $(\lambda,-\lambda)$.  The average 
spectral density of $D$ close to zero is nonstationary but universal 
and is of relevance for the physics of QCD at low energies.  It is 
determined by one of three so-called {\it chiral} Gaussian ensembles, 
where different ensembles correspond to different choices of the 
gauge group and the number of flavors\cite{jjmv}.  These ensembles 
have appeared in the context of disordered single-electron systems, 
too\cite{gade}.

In the present paper we introduce and analyse four additional Gaussian
random-matrix ensembles, which share many striking similarities with
the chiral ones but are demonstrably distinct.  The universality
classes they describe are realized in mesoscopic NS-systems, i.e. in
microstructures composed of a metallic part in contact with one or
several superconducting regions.  Just as in the classic Wigner-Dyson
case, the universality classes are distinguished by their behavior
under time reversal and rotation of the (electron's) spin.  The four
new classes together with the six known ones add up to a grand total
of ten.  We have reasons to believe that this exhausts the number of
possible universality classes in disordered single-particle systems 
and none else will be found.  (More precisely speaking, by 
universality classes we here mean infrared RG fixed points describing 
an ergodic limit.)  Some of our ideas were anticipated in 
\cite{oppermann,spm}.

\begin{figure}
\centerline{\psfig{figure=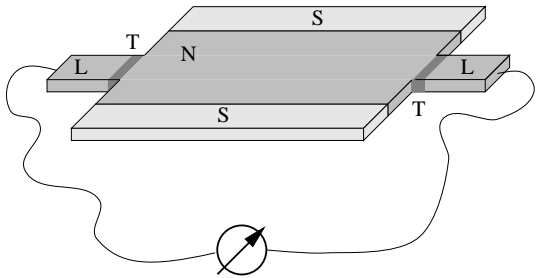,height=5.0cm}}
\caption{Metallic quantum dot (N) in contact with two
 superconducting regions (S).  The dot is separated from the
 leads (L) by a tunnel barrier (T).}
\label{fig1}
\end{figure}

The prototype of the kind of system we are going to study is depicted
in Fig.~\ref{fig1}.  A metallic (i.e. normal-conducting) quantum dot
is put in contact, via potential barriers, with two superconducting
regions.  Several leads are attached for the purpose of making current
and voltage measurements.  The metallic quantum dot may or may not be
disordered.  In the latter case we assume its geometric shape to be
such that the classical motion of a single electron inside it is
chaotic.  The quantum dot may be pierced by a magnetic flux of the
order of one or several flux quanta, and there may exist some impurity
atoms causing spin-orbit scattering.  The temperature is so low that
the electron's phase coherence length exceeds the size of the quantum
dot by far.

\narrowtext

The characteristic feature that distinguishes the above kind of
quantum dot from more conventional mesoscopic systems, is the
possibility for two electrons to tunnel through the potential barrier
at the NS-interface, thereby adding a Cooper pair to (or removing it
from) the superconducting condensate.  An equivalent statement in
single-particle language is that an electron incident on the
NS-interface may be retroreflected as a hole (and vice versa).  This
process of particle-hole conversion, which conserves energy, momentum
and spin but violates charge, is called {\it Andreev reflection}.  In
the semiclassical limit, Andreev reflections give rise to numerous
almost-periodic orbits whose action does not grow but remains of order
$\hbar$ as the length of the orbit increases\cite{az}.  The existence
of these orbits modifies the mean density of states (Weyl term) of the
quantum dot without leads: in general, an excitation gap opens and we
arrive at the ``boring'' situation where the vicinity of the chemical
potential is devoid of single-particle states.  However, by tuning the
phase difference of the order parameters of the two superconducting
regions to the special value $\pi$, we can make the gap close.  More
generally, we expect quasiparticle excitations to exist right at the
chemical potential whenever the phase shift incurred during Andreev
reflection vanishes on average over the NS-interfacial region.
Disorder or dynamically generated chaos mixes the states and creates a
universal spectral region close to the chemical potential.  Its width
is determined by the energy uncertainty which is caused by the
coupling of particles and holes by Andreev reflection.  It is this
very region and its consequences for the transport properties that we
are going to study in the present paper. 

The organization of the paper is as follows.  Mesoscopic
independent-quasiparticle systems are classified according to their
behavior under time reversal and spin rotations in
Sec.~\ref{sec:class}.  Having specified the required dynamical input
in Sec.~\ref{sec:dyn_input} we formulate the appropriate random-matrix
ensembles in Sec.~\ref{sec:rme}.  In Sec.~\ref{sec:spectrum} we
discuss the spectral statistics of an isolated system, using first the
Dyson-Mehta orthogonal polynomial method and then diagrammatic
perturbation theory.  The latter method easily extends to the
calculation of the transport properties of an open system.  In
Sec.~\ref{sec:wl} we work out the weak localization correction to the
average conductance and in Sec.~\ref{sec:ucf} the universal
conductance fluctuations.  Our conclusions are presented in
Sec.~\ref{sec:conc}. 

\section{Symmetry Classification}
\label{sec:class}

The treatment of this paper is based on the BCS Hamiltonian
in the Hartree-Fock-Boboliubov mean-field approximation:
       \begin{eqnarray}
       \hat H &=& \int d^d x \left( \sum_{\sigma,\tau=\uparrow,
       \downarrow} \psi_\sigma^\dagger h_{\sigma\tau} \psi_\tau
       + \Delta \psi_\uparrow^\dagger \psi_\downarrow^\dagger
       + \Delta^* \psi_\downarrow \psi_\uparrow\right), \nonumber \\
       h &=& ({\bf p}-e{\bf A})^2/2m + V + {\bf U}_{\rm SO}
       \cdot \mbox{\boldmath$\sigma$} \times ({\bf p}-e{\bf A})-\mu.
       \nonumber
       \end{eqnarray}
Here $V(x)$ is a scalar potential which may have a random component,
and $\Delta(x)$ is the pairing field. The presence of a magnetic
vector potential ${\bf A}(x)$ breaks time reversal symmetry while the
spin-orbit field ${\bf U}_{\rm SO}(x)$ breaks invariance under
rotations of the electron's spin. $\mu$ is the chemical potential.

The second-quantized Hamiltonian $\hat H$ can be rewritten in an
equivalent first-quantized form by the Boboliubov-deGennes (BdG)
formalism.  For our purposes it is convenient to introduce some
generic orthonormal basis of single-electron states $|\alpha\rangle$,
where $\alpha$ is a multi-index that combines the orbital and spin
quantum numbers of the electron. If $N$ is the number of orbital
states used, $\alpha$ runs from $1$ to $2N$. Let $c_\alpha^\dagger$
and $c_\beta$ be the usual creation and annihilation operators obeying
the canonical anticommutation relations $c_\alpha^\dagger
c_\beta^{\vphantom {\dagger}} + c_\beta^{\vphantom{\dagger}}
c_\alpha^\dagger = \delta_{\alpha\beta}$.  The Hamiltonian $\hat H$
can be written
      \[
      \hat H = \sum_{\alpha\beta} \left( h_{\alpha\beta} 
      c_\alpha^\dagger c_\beta^{\vphantom{\dagger}} + 
      {\textstyle{1\over 2}} \Delta_{\alpha\beta} 
      c_\alpha^\dagger c_\beta^\dagger + {\textstyle{1\over 2}} 
      \Delta_{\alpha\beta}^* c_\beta c_\alpha \right) .
      \]
Hermiticity requires $h_{\alpha\beta} = h_{\beta\alpha}^*$, and the
matrix elements $\Delta_{\alpha\beta}$ must be antisymmetric by Fermi
statistics: $\Delta_{\alpha\beta} = - \Delta_{\beta\alpha}$.  Now
we write $\hat H$ in the form ``row multiplies matrix multiplies
column'':
        \begin{equation}
        \hat H = {1\over 2} \left( {\bf c}^\dagger
        \quad {\bf c} \right)
        \pmatrix{h &\Delta\cr -\Delta^* &-h^{\rm T}\cr} 
        \pmatrix{{\bf c}\cr {\bf c}^\dagger\cr}  + {\rm const}.
        \label{rmmmc}
        \end{equation}
In this way every Hamiltonian $\hat H$ is uniquely assigned to a 
$4N\times 4N$-matrix ${\cal H}$,
      \begin{equation}
      {\cal H} = \pmatrix{h &\Delta\cr -\Delta^* &-h^{\rm T}\cr} .
      \label{BdG}
      \end{equation}
The eigenvalue problem for ${\cal H}$ is known as the
Bogoliubov-deGennes equations.  We refer to ${\cal H}$ as the
``BdG-Hamiltonian'' for short.

The first-quantized Hamiltonian ${\cal H}$ acts in an enlarged space,
namely the tensor product of the physical space ${\bf C}^{2N}$
(orbitals and spin) with an extra degree of freedom ${\bf C}^2$,
which we call the ``particle-hole space''.  Note however that the
``particles'' and ``holes'' of the BdG-formalism are not the particle
and hole states of a degenerate Fermi gas.  Indeed, the matrix $h$
already acts on {\it all} of the single-electron states, which have
energies {\it either above or below} the chemical potential.  The
BdG-hole states acted upon by $-h^{\rm T}$ are identical (and in this
sense redundant, or unphysical) copies of the BdG-particle states
acted upon by $h$.  They are introduced for the convenience of
treating the pairing field within the formalism of first quantization.

The aim of the current section is to classify systems according to
their symmetries.  Using the BdG-formalism we will show that the
presence or absence of time-reversal and/or spin-rotation invariance
leads to four distinct symmetry classes.  The situation thus is
different from the well-known Wigner-Dyson scenario where only three
distinct classes exist.

The discussion of Secs.~\ref{sec:scD}-\ref{sec:sCI} uses some basic
facts from the theory of Lie algebras and symmetric spaces and is
somewhat technical.  The casual reader may wish to skip these
subsections and proceed directly to Table 1 given at the end of
Sec.~\ref{sec:sCI}.  A brief summary of the symmetries of ${\cal H}$
for each class is provided also at the beginning of
Sec.~\ref{sec:rme}.

\noindent\subsection{Symmetry Class $D$}
\label{sec:scD}

We start by considering systems with the least degree of symmetry,
i.e. systems with neither time-reversal symmetry nor spin-rotation
invariance.  In this case the matrices $h$ and $\Delta$ in general
have no symmetry properties beyond those stated above, namely
hermiticity of $h$ and skew-symmetry of $\Delta$.  Because the set of
hermitian matrices does not close under commutation whereas the
antihermitian ones do, we prefer to work with $X := i{\cal H}$ rather
than ${\cal H}$ in the current section.  In terms of $X$, the
conditions $h = h^\dagger$, $\Delta = -\Delta^{\rm T}$ can be
presented summarily in the form
      \begin{eqnarray}
      -X^\dagger &=& X = - \Sigma_x X^{\rm T} \Sigma_x, 
      \label{orthogonal}\\
      \Sigma_x &=& \pmatrix{0 &{\bf 1}_{2N}\cr {\bf 1}_{2N}
      &0\cr} = \sigma_x \otimes {\bf 1}_{2N} .\nonumber
      \end{eqnarray}
If two matrices $X,Y$ satisfy these equations, then so does their
commutator $[X,Y]$. Hence, we may view $X$ as an element of some Lie
algebra. To identify this Lie algebra we conjugate by $X \mapsto
\tilde X = U_0 X U_0^{-1}$ where $U_0 = {\textstyle{1\over \sqrt{2}}}
\left(\begin{array}{cc} 1&1\\ i&i
\end{array}\right) \otimes {\bf 1}_{2N}$.
Equations (\ref{orthogonal}) then take the form $-\tilde X^\dagger =
\tilde X = - \tilde X^{\rm T}$ or, equivalently, $\tilde X = \tilde
X^* = -\tilde X^{\rm T}$.  This shows that (\ref{orthogonal}) fixes a
${\rm so}(4N)$-algebra, i.e. a Lie algebra isomorphic to the real
antisymmetric $4N\times 4N$-matrices. Since ${\rm so}(4N) \equiv 
D_{2N}$ in Cartan's notation, we denote the present symmetry class 
by the symbol ``$D$''.

Being a Lie algebra element, $X$ can be brought into diagonal form by
$X \mapsto \Omega = gXg^{-1}$ where $g$ is an element of the
corresponding Lie group which is isomorphic to ${\rm SO}(4N)$ and is
defined by
      \begin{equation}
      {g^{-1}}^\dagger = g = \Sigma_x {g^{-1}}^{\rm T} \Sigma_x .
      \label{Orthogonal}
      \end{equation}
The conditions (\ref{orthogonal}) imply $\Omega = {\rm diag}\;\left(
  i\omega,-i\omega\right) = \sigma_z \otimes i\omega$ where $\omega =
{\rm diag} (\omega_1, \omega_2, ..., \omega_{2N})$ with real
$\omega_i$.  The conditions for the canonical anticommutation
relations to be invariant under a transformation
      \[
      \pmatrix{{\bf c}^{\vphantom{\dagger}}\cr {\bf c}^\dagger\cr} 
      \mapsto g \pmatrix{{\bf c}^{\vphantom{\dagger}}\cr 
      {\bf c}^\dagger\cr} =:
      \pmatrix{\mbox{\boldmath$\gamma$}^{\vphantom{\dagger}}\cr 
      \mbox{\boldmath$\gamma$}^\dagger\cr}
      \]
can be shown\cite{balian} to coincide with (\ref{Orthogonal}).
Thus, inserting $X = g^{-1} \Omega g$ into (\ref{rmmmc}) we obtain
      \[
      \hat H = {1\over 2} \sum_\lambda \omega_\lambda
      \left( \gamma_\lambda^\dagger \gamma_{\lambda}^{\vphantom
      {\dagger}} - \gamma_\lambda^{\vphantom{\dagger}}
      \gamma_\lambda^\dagger \right)
      \]
with $\gamma_\alpha^\dagger \gamma_\beta^{\vphantom{\dagger}} +
\gamma_\beta^{\vphantom{\dagger}} \gamma_\alpha^\dagger = 
\delta_{\alpha\beta}$. The frequencies $\omega_\lambda$ may be
positive or negative. The BCS ground state is defined by
demanding $\gamma_\lambda |{\rm BCS}\rangle = 0$ for
$\omega_\lambda > 0$ and $\gamma_\lambda^\dagger |{\rm BCS}
\rangle = 0$ for $\omega_\lambda < 0$. The normal-ordered
Hamiltonian
      \[
      :\hat H: \ = 
      \sum_{\omega_\lambda > 0} |\omega_\lambda|
      \gamma_\lambda^\dagger \gamma_\lambda^{\vphantom{\dagger}}
      + \sum_{\omega_\lambda < 0} |\omega_\lambda|
      \gamma_\lambda^{\vphantom{\dagger}} \gamma_\lambda^\dagger
      \]
is always positive. 

\subsection{Symmetry Class $C$}

We now consider systems without time-reversal symmetry but {\it with}
spin-rotation invariance. We again use the unique representation of a
second-quantized BCS mean-field Hamiltonian $\hat H$ by ($i$ times) a
BdG-Hamiltonian $X = i{\cal H}$.

We write the particle-hole decomposition of $X$ as $X = \left(
\begin{array}{cc}A&B\\ C&D\end{array}\right)$ or, 
in tensor-product notation,      
      \[
      X = E_{\rm pp} \otimes A + E_{\rm ph} \otimes B + E_{\rm hp}
      \otimes C + E_{\rm hh} \otimes D .
      \]
The condition $X = - \Sigma_x X^{\rm T} \Sigma_x$ means $B = - B^{\rm
  T}$, $C = -C^{\rm T}$ and $D = -A^{\rm T}$. Antihermiticity requires
$A = -A^\dagger$ and $C = -B^\dagger$.

The generators of spin rotations, $J_k$ $(k=x,y,z)$, are represented
on particle-hole space by $J_k = (E_{\rm pp}\otimes\sigma_k - E_{\rm
  hh} \otimes\sigma_k^{\rm T})\otimes{\bf 1}_{N}$.  Spin-rotation
invariance of the Hamiltonian requires that $X$ and $J_k$ commute.
This condition is easily seen to constrain $A,B,C$ to be of the form
$A = {\bf 1}_2 \otimes a$, $B = i\sigma_y \otimes b$ and $C =
-i\sigma_y \otimes c$ or, in matrix presentation,
      \[
      X = \pmatrix{a &0 &0 &b\cr 0 &a &-b &0\cr
      0 &-c &-a^{\rm T} &0\cr c &0 &0 &-a^{\rm T}\cr} .
      \]
We see that $X$ decomposes into two commuting subblocks.  One
corresponds to spin-up particles and spin-down holes, and the other to
spin-down particles and spin-up holes.  Because the subblocks are
related by an algebra homomorphism ($b \mapsto -b$, $c \mapsto -c$) it
is sufficient to focus on one of them, say
      \[
      X_\uparrow = \pmatrix{a &b\cr c &-a^{\rm T}\cr}
      \]
and account for spin degeneracy by inserting factors of two whenever
needed. We drop the subscript and write $X$ for $X_\uparrow$.

Since $B = -B^{\rm T}$, the equation $B = i\sigma_y \otimes b$ implies
$b = +b^{\rm T}$.  Similarly, we deduce $c = + c^{\rm T}$.
Antihermiticity requires $a = -a^\dagger$ and $c = -b^\dagger$.  All
these conditions are summarized by
      \begin{equation}
      -X^\dagger = X = -\Sigma_y X^{\rm T} \Sigma_y,
      \quad \Sigma_y  = \sigma_y \otimes {\bf 1}_N .
      \label{symplectic}
      \end{equation}
This is the defining equation of the symplectic Lie algebra ${\rm
  sp}(2N)$. Thus $X = i{\cal H}$ is an element of ${\rm sp}(2N)$.
Since ${\rm sp}(2N) \equiv C_N$ in Cartan's notation, we denote the
present symmetry class by ``$C$''.

The second-quantized Hamiltonian associated with $X = X_\uparrow =
\left(\begin{array}{cc}a&b\\c&-a^{\rm T}\end{array}\right)$ is
      \[
      \hat H_\uparrow = -{i\over 2} \sum_{m,n} (c_{m\uparrow}^\dagger
      \quad c_{m\downarrow}) \pmatrix{a_{mn} &b_{mn}\cr c_{mn}
      &-a_{nm}\cr} \pmatrix{c_{n\uparrow}\cr c_{n\downarrow}^\dagger}.
      \]
As before, we can diagonalize $X$ by $X = g^{-1} \Omega g$ where
$\Omega = \sigma_z\otimes i\omega$ and $\omega = {\rm
  diag}(\omega_1,...,\omega_M)$, and $g$ now satisfies
${g^{-1}}^\dagger = g = \Sigma_y {g^{-1}}^{\rm T} \Sigma_y$, i.e. $g
\in {\rm Sp}(2N)$. The transformation
      \[
      \pmatrix{{\bf c}_\uparrow\cr {\bf c}_\downarrow^\dagger\cr} 
      \mapsto g \pmatrix{{\bf c}_\uparrow\cr 
      {\bf c}_\downarrow^\dagger\cr} =:
      \pmatrix{\mbox{\boldmath$\gamma$}_\uparrow\cr 
      \mbox{\boldmath$\gamma$}_\downarrow^\dagger\cr}
      \]
diagonalizes the Hamiltonian:
      \begin{eqnarray}
        \hat H &=& \hat H_\uparrow + \hat H_\downarrow \nonumber \\ 
        &=& {1\over 2} \sum_\lambda \omega_\lambda \left(
          \gamma_{\lambda\uparrow}^\dagger
          \gamma_{\lambda\uparrow}^{\vphantom{\dagger}} +
          \gamma_{\lambda\downarrow}^\dagger
          \gamma_{\lambda\downarrow}^{\vphantom{\dagger}} -
          \gamma_{\lambda\uparrow}^{\vphantom{\dagger}}
          \gamma_{\lambda\uparrow}^\dagger -
          \gamma_{\lambda\downarrow}^{\vphantom{\dagger}}
          \gamma_{\lambda\downarrow}^\dagger \right). \nonumber
      \end{eqnarray}
Because $g \in {\rm Sp}(2N) \subset {\rm U}(2N)$ is a unitary 
matrix, the canonical anticommutation relations are preserved by the
transformation from $({\bf c},{\bf c}^\dagger)$ to 
$(\mbox{\boldmath$\gamma$},\mbox{\boldmath$\gamma$}^\dagger)$.
Every quasiparticle level has a trivial degeneracy due to
spin.

\subsection{Symmetry Class $D$III}

We next consider systems with time-reversal symmetry but without 
spin-rotation invariance.  Recall that the conditions for symmetry 
class $D$, $-X^\dagger = X = -\Sigma_x X^{\rm T} \Sigma_x$ with 
$\Sigma_x = \sigma_x \otimes {\bf 1}_2 \otimes {\bf 1}_N$, fix a 
${\rm so}(4N)$-algebra. The time-reversal operator ${\cal T}$ acts 
on the BdG-Hamiltonian by
      \[
      {\cal T} : {\cal H}\mapsto \tau {\cal H}^* \tau^{-1}
      \]
where $\tau = {\bf 1}_2 \otimes i\sigma_y \otimes {\bf 1}_N$.
Using $X = i{\cal H}$ and $X^* = -X^{\rm T}$ we get the
action of ${\cal T}$ on $X$, ${\cal T} : X \mapsto \tau X^{\rm T}
\tau^{-1}$. Thus, for a time-reversal invariant system, $X$
is subject to the additional constraint $X = +\tau X^{\rm T}
\tau^{-1}$. We denote the set of solutions in ${\rm so}(4N)$ 
of this condition by ${\cal P}$. While ${\cal P}$ does not
close under commutation and therefore does not form a 
subalgebra of ${\rm so}(4N)$, the solution set, ${\cal K}$,
of the complementary condition $Y = -\tau Y^{\rm T} \tau^{-1}$
does. Therefore, we may describe ${\cal P}$ as the complement
of a subalgebra ${\cal K}$ in ${\rm so}(4N)$. In formulas,
${\rm so}(4N) = {\cal K} + {\cal P}$. We are now going to
identify ${\cal K}$.

The equations for ${\cal K}$ can be rewritten
      \[
      -Y^\dagger = Y = -\Sigma_x Y^{\rm T} \Sigma_x
      = - (\Sigma_x\tau) Y (\Sigma_x \tau)^{-1} .
      \]
Let $U_0$ be the unitary matrix given in particle-hole
and spin decomposition by
      \[
      U_0 = {1\over \sqrt{2}}
      \pmatrix{{\bf 1}_2 &i\sigma_y\cr \sigma_y &-i{\bf 1}_2\cr}
      \otimes {\bf 1}_N .
      \]
Conjugation by $U_0$, $Y \mapsto \tilde Y = U_0^{-1} Y U_0$,
takes the equations for ${\cal K}$ into
      \[
      -{\tilde Y}^\dagger = \tilde Y = 
      -\Sigma_x {\tilde Y}^{\rm T} \Sigma_x
      = - \Sigma_z {\tilde Y} \Sigma_z .
      \]
The solutions of the latter are of the form $\tilde Y =
{\rm diag}\;\left(Z,-Z^{\rm T}\right)$ with $Z$ an
antihermitian $2N\times 2N$-matrix. We now recognize ${\cal K}$ as
being isomorphic to the Lie algebra of antihermitian $2N\times
2N$-matrices, or ${\cal K} \simeq {\rm u}(2N)$. Thus, the space
${\cal P}$ of BdG-Hamiltonians $X = i{\cal H}$ is obtained from
${\rm so}(4N)$ by removing a ${\rm u}(2N)$-subalgebra.  Because
${\cal P}$ is the difference of two Lie algebras ${\rm so}(4N)$ and
${\rm u}(2N)$, it can be interpreted as the tangent space of
the quotient ${\rm SO}(4N)/{\rm U}(2N)$ of the corresponding
Lie groups, which is a symmetric space of type $D$III in Cartan's 
notation; hence the name $D$III for the present symmetry class.

{}From the general theory of symmetric spaces\cite{helgason} we know
that an element $X \in {\cal P}$ can be diagonalized by a
transformation $X \mapsto k^{-1} X k$ with $k \in \exp{\cal K} = {\rm
U}(2N)$. Time-reversal symmetry causes every eigenvalue to be doubly
degenerate by Kramers' theorem. 

\subsection{Symmetry Class $C$I}
\label{sec:sCI}

Finally, we turn to systems with both time-reversal symmetry and
spin-rotation invariance. Recall that spin-rotation invariance causes
the BdG-Hamiltonian $X = i{\cal H}_\uparrow$ to obey the relations
$-X^\dagger = X = -\Sigma_y X^{\rm T} \Sigma_y$, which define the
symplectic Lie algebra ${\rm sp}(2N)$.  Because of the restriction to
a single spin sector, the action of the time reversal operator
simplifies to ${\cal T} : X \mapsto X^{\rm T}$. Thus, time-reversal
symmetry constrains $X$ to be symmetric. Let ${\cal K}$ now denote the
subalgebra of antisymmetric matrices in ${\rm sp}(2N)$.  Then $X$,
being symmetric, lies in the complementary set ${\cal P}$ defined by
${\rm sp}(2N) = {\cal K} + {\cal P}$. We claim that ${\cal K}$ is
isomorphic to the unitary Lie algebra ${\rm u}(N)$.  To prove this, we
observe that the solutions $Y \in {\cal K}$ of $-Y^\dagger = Y =
-\Sigma_y Y^{\rm T} \Sigma_y = Y^{\rm T}$ have the form ${\bf 1}_2
\otimes {\rm Re}A + i\sigma_y \otimes {\rm Im}A$ where $A$ is an
arbitrary antihermitian $N\times N$-matrix, i.e.  $A\in {\rm u}(N)$.
The embedding ${\rm u}(N) \mapsto {\rm sp}(2N)$ by $A \mapsto {\bf
1}_2 \otimes {\rm Re}A + i\sigma_y \otimes {\rm Im}A$ is easily seen
to be a homomorphism of Lie algebras.  Therefore ${\cal K} \simeq {\rm
u}(N)$ as claimed.  The linear complement ${\cal P}$ of ${\rm u}(N)$
in ${\rm sp}(2N)$ can be regarded as the tangent space of ${\rm
Sp}(2N)/{\rm U}(N)$, which is a compact symmetric space of type $C$I
according to Cartan's list. 

For the benefit of the casual reader the various symmetry classes and
the names by which they are referred to in the present paper,
are summarized in Table 1.
\narrowtext
\begin{center}
\begin{tabular}{|c||c|c|c|}\hline
Class & {time-rev.} 
             & {spin-rot.} 
             & sym. space\\ \hline
$D$   & no& no&SO(4$N$)\\
$C$   & no&yes&Sp(2$N$)\\
$D$III&yes& no&SO(4$N$)/U(2$N$)\\
$C$I  &yes&yes&Sp(2$N$)/U($N$)\\ \hline
\end{tabular}\\
\bigskip
Table 1
\end{center}

\subsection{Is the symmetry (\ref{orthogonal}) wiped out by 
Coulomb effects?} 
\label{sec:coulomb}

The symmetry (\ref{orthogonal}) is central to our approach.  Just how
robust is it? 

The relations (\ref{orthogonal}) follow from the well-known
mathematical fact\cite{balian} that the set of all bilinear
combinations of the fermion creation and annilihation operators is
isomorphic to an orthogonal Lie algebra.  Put differently, the
symmetry (\ref{orthogonal}) requires no more than the fermionic nature
of the electron and the use of the Hartree-Fock-Bogoliubov mean-field
approximation, allowing us to express the Hamiltonian in terms of
bilinears of the creation and annihilation operators.  Alternatively,
we could say that (\ref{orthogonal}) is an exact symmetry whenever the
system can be described in terms of independent Boboliubov
quasiparticles.

What happens when we add a Coulomb charging energy to the Hamiltonian?
The relative minus sign between the particle-particle and hole-hole
blocks of ${\cal H}$, Eq.~(\ref{BdG}), tells us that, if the creation
of an electron in a given state costs a certain amount of energy, then
the creation of a hole (removal of an electron) in this state should
release exactly the same amount.  The Coulomb interaction, however,
does not conform to that principle.  When a charge is added to a
charge-neutral system, say, it makes no difference whether this charge
is a particle or a hole, the electrostatic energy cost is positive in
both cases.  Therefore, the Coulomb charging energy (as well as other
perturbations that do not fit into the independent-quasiparticle
framework) violates the symmetry (\ref{orthogonal}).  More precisely
speaking, we expect the independent-quasiparticle approximation to be
adequate for describing the short-time physics, but at sufficiently
long times Coulomb effects must become visible and, in particular,
they will cut off the particle-hole modes we are going to study in the
present paper.  Whether the cutoff time can be long enough for the
consequences of the symmetry (\ref{orthogonal}) to be observable, is a
tough quantitative question for theory, which cannot be answered
without an understanding of screening in open and finite metallic
systems.  Fortunately, the question has already been answered in the
affirmative by experiment.  Over the last years a number of novel
mesoscopic NS-phenomena has been observed, the most prominent of which
is the dramatic enhancement\cite{kastalsky} of the differential
conductance at zero bias in NS-geometries with a high potential
barrier separating the normal-conducting and superconducting regions.
This phenomenon has been explained\cite{vanWees} by a mechanism called
``coherent Andreev reflection'' or ``reflectionless
tunneling''\cite{beenakker}, which is the result of constructive
interference between semiclassical paths with one Andreev reflection
and a variable number of normal reflections.  In order for such an
interference to take place, the dynamical phases of a particle and a
hole traversing the same path in opposite directions must cancel each
other.  It is precisely the symmetry (\ref{orthogonal}) in combination
with the extra symmetries defining class $C$I, that guarantee the
necessary phase relation between particles and holes to hold.  We
conclude that there exists convincing experimental evidence that the
symmetry (\ref{orthogonal}) is not wiped out by the Coulomb
interaction but leads to observable consequences.  In the remainder of
this paper we are going to ignore Coulomb effects.

\section{Dynamical Input}
\label{sec:dyn_input}

The classification of Sec.~\ref{sec:class} refers only to symmetry
and thus is very general.  To go further, we make two dynamical
assumptions.

When an electron is retroreflected from the NS-interface as a hole,
its wavefunction acquires a phase shift which is determined by the
phase of the superconducting order parameter.  Our first assumption is
that this phase shift, here called the ``Andreev phase shift'' for
short, {\it vanishes on average over the NS-interfacial region}.  To
appreciate what such an assumption implies, let us look at a few
examples.  Consider first an SNS-system consisting of an infinite slab
of normal metal sandwiched between two superconducting slabs $S_1$ and
$S_2$.  The pairing interaction causes the existence of an excitation
gap in each of the superconducting regions.  We now ask how the
presence of the normal-conducting slab affects the excitation spectrum
of the combined SNS-system at the chemical potential.  The answer to
this question was first given in\cite{andreev2,kulik} and it is
essentially as follows.  In the clean limit, the BdG-Hamiltonian is
separable and we can get a qualitative understanding of the quantal
energy spectrum by the method of semiclassical quantization.  For
simplicity we assume that all reflections at the NS-interface are
Andreev.  Every periodic classical motion then is some multiple of a
primitive periodic orbit where an electron moves back and forth
between the superconducting regions and is Andreev reflected at each
interface.  If $k_{\rm p}$ ($k_{\rm h}$) are the wave numbers of the
particle (hole) normal to the slabs and $L$ is the thickness of the
normal-conducting region, the Bohr-Sommerfeld quantization rule
applied to this periodic motion reads
        \begin{equation}
        \pm (k_{\rm p}-k_{\rm h})L + \pi + \varphi_1  - 
        \varphi_2 = 2\pi n \qquad (n\in{\bf Z}).
        \label{BS}
        \end{equation}
Here $\pi+\varphi_1-\varphi_2$ is the phase accumulated by the two
Andreev reflections if $\varphi_1$ and $\varphi_2$ are the phases of
the superconducting order parameter in the regions $S_1$ and $S_2$.
For an electron with energy equal to the Fermi energy, $k_{\rm p}$
equals $k_{\rm h}$, so the first term on the left-hand side vanishes.
{}From the resulting equation $\varphi_1-\varphi_2 = (2n-1)\pi$ we see
that the quantization condition can be fulfilled only when $\varphi_1$
and $\varphi_2$ differ by an odd multiple of $\pi$.  In other words,
for $\cos(\varphi_1-\varphi_2) \not= -1$, which includes the
homogeneous case $\varphi_1 = \varphi_2$, we expect a gap in the
excitation spectrum not only in the superconductor but also in the
combined SNS-system.  On the other hand, for $\cos(\varphi_1 -
\varphi_2) \to -1$ the gap closes and quasiparticle excitations exist 
all the way down to zero energy.  The latter situation is special in 
that the Andreev phase shift vanishes on average over the 
two NS-interfaces in that case. 

The above argument applies to the extreme limit of a clean system
which clearly is unrealistic. What can we say about the effects of
disorder?  A generic random potential destroys separability and makes
Bohr-Sommerfeld quantization inapplicable.  The general case therefore
needs to be studied with the help of a computer, or by using the
random-matrix theory that will be developed in the remainder of our
paper.  What is easy to treat analytically is the case of a slowly
varying random potential depending only on the coordinate, $z$, of the
direction perpendicular to the slabs.  In this case the quantization
rule (\ref{BS}) remains valid if we replace the expression $(k_{\rm
p}-k_{\rm h})L$ by the integral $\int_0^L \Bigl(k_{\rm p}(z) - k_{\rm
h}(z) \Bigr) dz$ where $k_{\rm p}(z) = [2m ( \mu +
\varepsilon - V(z)]^{1/2}$, $k_{\rm h}(z) = [2m ( \mu -
\varepsilon - V(z)]^{1/2}$, and $E = \mu +\varepsilon$ is
the total energy of the electron.  Since $k_{\rm p}(z)
= k_{\rm h}(z)$ for $\varepsilon = 0$, our conclusions are
the same as before: there is a gap for $\cos(\varphi_1 - 
\varphi_2) \not= -1$, and the gap closes as $\cos(\varphi_1
-\varphi_2) \to -1$. 

Another instructive example is provided by the vortex solution for a
clean type-II superconductor.  The phase of the superconducting order
parameter uniformly winds once around the unit circle as we go once
around the vortex core.  For this reason, the pairing field
experienced by normal excitations bound to the vortex core vanishes on
average over the vortex.  Because the vortex solution breaks
translational symmetry, there must exist some RPA (or vibrational)
zero modes of the vortex.  These zero modes are the Goldstone modes
associated with the spontaneous breaking of translational symmetry by
the localized vortex solution.  It follows that, if the RPA
correlation energy vanishes (is small), i.e. if the excitation
energies are given by (are approximately given by) sums of two
quasiparticle energies, there must exist quasiparticle excitations
with vanishing (small) energy.  In contrast, for a piece of
cylindrically shaped normal metal immersed in a superconducting
environment (``columnar defect'') there is no general reason why we
should expect quasiparticle excitations with low energy. 

These two examples, the SNS slab geometry and the vortex, lend support
to the plausible expectation that a pairing field which is locally
nonzero but whose mean phase $\langle e^{i\phi} \rangle$ vanishes
in a suitably defined sense, is ineffective at creating a genuine gap
in the density of states near the chemical potential.  This then is
the motivation behind the above requirement that the Andreev phase
shift should vanish on average over the NS-interfacial region: it
ensures that the gap closes and quasiparticles can exist right at the
chemical potential.

Our second main input is the assumption that {\it the classical
dynamics in the normal-conducting (N-) region be chaotic}.  The
presence of a sufficient amount of disorder will always guarantee this
assumption to be justified.  For a ballistic system, chaotic dynamics
is achieved by choosing for the boundary of the N-region some surface
that causes a typical classical trajectory to be unstable.  Chaoticity
of the classical motion means that the long-time behavior of the
system is unpredictable; in particular, the phase shifts acquired by
Andreev reflection along a typical semiclassical trajectory form a
random sequence.  This randomness will allow us to model the pairing
field by a stochastic variable.  Note that the spatial constancy of
the magnitude of the pairing field in the bulk of the superconductor
is an irrelevant feature for our purposes. If we switch from
coordinate representation to a generic basis of single-particle
states, say the eigenbases of $h$ and $-h^{\rm T}$, both the phase and
the magnitude of $\Delta$ will fluctuate and be distributed around
zero. 

Consider now an isolated finite system, so that the Boboliubov
quasiparticle spectrum is discrete.  According to our above arguments,
we expect the existence of levels close to the chemical potential in
the pure system under the conditions described.  The effect of
dynamically generated chaos and/or disorder will be to cause mixing of
these levels.  For the conventional N-system, such mixing is known to
lead to universal level statistics, depending only on symmetry.  (More
precisely, the level correlations are universal in the low-frequency
regime corresponding to the long-time or ergodic limit.) For the case
of disordered metallic granules, the level correlations were
calculated by Efetov\cite{efetov}.  His results show that the level
statistics is Wigner-Dyson, i.e.  identical to that of an ensemble of
random matrices with uncorrelated Gaussian distributed matrix
elements.  In the NS-systems considered in the present paper, new
features appear at low excitation energy, owing to the coupling of
particles and holes by the process of Andreev reflection at the
NS-interface.  As was shown in Sec.~\ref{sec:class}, the presence of
the pairing field $\Delta$ leads to novel symmetries.  We therefore
expect new types of universal level statistics to emerge in such
systems.  The new type of correlation will extend over an energy range
set by the energy uncertainty due to the action of the pairing field
(or Andreev reflection).  The goal of our paper is to give a
quantitative description of precisely these correlations and their
effect on the transport properties.  To reach this goal we may follow
two different routes.  The first and more comprehensive one is to
generalize Efetov's analysis, i.e.  to construct an effective field
theory of the nonlinear $\sigma$ model type and solve the field theory
in the zero-dimensional limit corresponding to the universal regime.
Such an approach yields not only the universal behavior but also the
crossover to the short-time regimes.  Since our interest is in the
universal limit, there exists also another option.  Armed by the
experience gained from the study of the N-system, we may replace the
BdG-Hamiltonian (\ref{BdG}) by an ensemble of random matrices with
maximum entropy, paying attention only to the symmetries under time
reversal and spin rotation.  While the field-theoretic method is more
versatile, the random-matrix or maximum-entropy approach has the great
advantage of being much simpler technically.  For this reason we have
chosen to follow the latter in the present paper. 

To maximize the entropy of the random-matrix ensemble we will take the
matrix elements of the BdG-Hamiltonian to be normally distributed and
statistically independent.  All matrix elements will be chosen to have
{\it zero mean}.  For the off-diagonal blocks of the BdG-Hamiltonian,
this choice corresponds to our assumption that the spatial average of
the Andreev phase shift vanishes on average.  In general, we would
need to distinguish between the strength of fluctuation of $h$ and
$\Delta$.  However, at low energy, i.e. within the energy window
defined by the uncertainty due to Andreev reflection, this distinction
turns out to be irrelevant and we may take the strengths to be equal.
The resulting random-matrix ensemble depends on two parameters only.
These are the strengths of the perturbations that break time-reversal
and spin-rotation invariance and are responsible for the crossover
between universality classes. 

To finish off this orientational section, we mention another realm of
application of the above random-matrix ideas.  Consider an array of
superconducting grains or islands embedded in a metallic
(non-superconducting) host.  The grains are disordered and/or of
irregular shape, and they are mutually coupled by Josephson
tunnelling.  The array is exposed to a subcritical magnetic field
which penetrates the host but is ejected from the grains.  By tuning
the strength of the field we can frustrate the coupling between the
grains and drive the system into a spin-glass type of phase where
superconducting order exists locally but not globally.  Such a system
has been called a superconducting glass\cite{oppermann}.  Its prime
characteristic is that the pairing field, or superconducting order
parameter, continues to be nonzero on each grain but vanishes on
average over large scales.  The low-energy quasiparticle excitations
of such a system are predicted to be described by the random-matrix
model formulated below.  Because of the breaking of time-reversal
symmetry by the magnetic field, the relevant symmetry class is $C$.
The presence of spin-orbit interactions causes crossover to $D$.

\section{Random-Matrix Ensembles}
\label{sec:rme}

To prepare the formulation of the random-matrix ensembles, we
summarize the discussion of Sec.~\ref{sec:class} by presenting the
symmetries of the BdG-Hamiltonian ${\cal H}$ for each symmetry class
explicitly.  \smallskip

For systems where all symmetries are broken (class $D$) ${\cal H}$
satisfies ${\cal H} = - \Sigma_x {\cal H}^{\rm T} \Sigma_x$ with
$\Sigma_x = \sigma_x\otimes{\bf 1}_2\otimes {\bf 1}_N$. The block
decomposition
      \[
      {\cal H} = \pmatrix{A &B\cr B^\dagger &-A^{\rm T}\cr}
      \]
expresses the particle-hole structure of ${\cal H}$.  The
off-diagonal block $B$ is antisymmetric by Fermi statistics.  
Hermiticity of the Hamiltonian requires $A = A^\dagger$.

Class $D$III consists of the systems where time reversal is the only
good symmetry. For such systems ${\cal H}$ obeys the additional
relation ${\cal H} = \tau {\cal H}^{\rm T} \tau^{-1}$ with $\tau =
{\bf 1}_2 \otimes i\sigma_y \otimes {\bf 1}_N$. The decomposition of
${\cal H}$ according to particles and holes (outer block structure)
and spin (inner block structure) has the form
      \[
      {\cal H} = \pmatrix{
      a_{\uparrow\uparrow} &a_{\uparrow\downarrow}
      &b_{\uparrow\uparrow} &b_{\uparrow\downarrow}\cr
      a_{\downarrow\uparrow} &a_{\uparrow\uparrow}^{\rm T}
      &-b_{\uparrow\downarrow}^{\rm T} &b_{\downarrow\downarrow}\cr
      -b_{\downarrow\downarrow} &-b_{\uparrow\downarrow}^{\rm T}
      &-a_{\uparrow\uparrow}^{\rm T} &a_{\downarrow\uparrow}\cr
      b_{\uparrow\downarrow} &-b_{\uparrow\uparrow}
      &a_{\uparrow\downarrow} &-a_{\uparrow\uparrow}\cr}
      \]
with antisymmetric $a_{\uparrow\downarrow}$, $a_{\downarrow
\uparrow}$, $b_{\uparrow\uparrow}$ and $b_{\downarrow
\downarrow}$.  Hermiticity of ${\cal H}$ requires
$a_{\uparrow\uparrow} = a_{\uparrow\uparrow}^\dagger$,
$b_{\uparrow\downarrow} = b_{\uparrow\downarrow}^\dagger$,
$a_{\uparrow\downarrow} = a_{\downarrow\uparrow}^\dagger$, and
$b_{\downarrow\downarrow} = - b_{\uparrow\uparrow}^\dagger$.

For class $C$ spin is conserved while time-reversal symmetry is
broken.  In this case ${\cal H}$ commutes with the spin-rotation
generators $J_k = (E_{\rm pp} \otimes \sigma_k - E_{\rm hh}\otimes
\sigma_k^{\rm T}) \otimes {\bf 1}_N$ or, equivalently, ${\cal H}$
obeys ${\cal H} = J_k {\cal H} J_k$.  The particle-hole and spin
decomposition of ${\cal H}$ reads
      \[
      {\cal H} = \pmatrix{a &0 &0 &b\cr 0 &a &-b &0\cr 0 &-b^\dagger 
      &-a^{\rm T} &0\cr b^\dagger &0 &0 &-a^{\rm T}\cr}
      \]
with symmetric $b$.  Every level has a trivial twofold
degeneracy due to spin. Without loss of information we may focus on
the spin-up sector with reduced Hamiltonian
      \[
      {\cal H}_{\rm r} = \pmatrix{a &b\cr b^\dagger &-a^{\rm T}\cr}.
      \]
Hermiticity requires $a = a^\dagger$.

In class $C$I both spin rotations and time reversal are good
symmetries. The BdG-Hamiltonian satisfies ${\cal H} = \tau {\cal
H}^{\rm T} \tau^{-1} = J_k {\cal H} J_k$, and is constrained by these
symmetries to be of the form
      \[
      {\cal H} = \pmatrix{a &0 &0 &b\cr 0 &a &-b &0\cr
      0 &-b &-a^{\rm T} &0\cr b &0 &0 &-a^{\rm T}\cr}
      \]
with symmetric $a$ and $b$.  Hermiticity then implies that $a$ and $b$
are real matrices. 

Now recall the dynamical conditions formulated in
Sec.~\ref{sec:dyn_input}.  By assumption, the classical dynamics in
the N-system is chaotic and the Andreev phase shift vanishes on
average over the NS-interfacial region.  We therefore may replace the
BdG-Hamiltonian by a random matrix (of the appropriate symmetry) with
matrix elements that have zero mean.  The principle of least
information, or maximum entropy, then leads us to postulate a
random-matrix ensemble with a Gaussian probability distribution
      \begin{equation}
      \exp \left( - {\rm Tr} {\cal H}^2 / 2v^2 \right)
      d {\cal H}
      \label{gaussian}
      \end{equation}
for each symmetry class. Here $d{\cal H}$ denotes a Euclidean measure
on the linear space of BdG-Hamiltonians with metric ${\rm Tr}(d{\cal
H})^2$. 

More generally, we can formulate a two-parameter family of Gaussian
random-matrix ensembles which interpolates between all four symmetry
classes. Because a Gaussian distribution (with zero mean) is
completely specified by its second moment, it is sufficient to
describe the correlation function $\langle{\rm Tr}{\cal AH} \times
{\rm Tr}{\cal BH}\rangle$ for two arbitrary sources ${\cal A}$ and
${\cal B}$. The correlation law we choose is
      \begin{eqnarray}
      &&\langle{\rm Tr}{\cal AH}\times
      {\rm Tr}{\cal BH}\rangle / v^2 \nonumber \\
      = &&{\rm Tr}({\cal A}-\Sigma_x{\cal A}^{\rm T}\Sigma_x)
      \Bigl[ {\cal B} + (1-\epsilon_t) \tau{\cal B}^{\rm T}\tau^{-1} 
      \nonumber \\
      &&\hskip 3.5cm + (1-\epsilon_s) \sum_k J_k {\cal B} J_k
      \label{correlator} \\ &&\hskip 1.5cm  
      + (1-\epsilon_s)(1-\epsilon_t) \sum_k J_k \tau B^{\rm T}
      \tau^{-1} J_k \Bigr].
      \nonumber
      \end{eqnarray} 
For $\epsilon_s = \epsilon_t = 0$ the correlation law is invariant
under both reversal of time and rotation of spin.  This is the
symmetry class $C$I. A nonzero value of $\epsilon_t$ breaks 
time-reversal symmetry.  Therefore, by increasing $\epsilon_t$ we 
cross over to class $C$. A nonzero value of $\epsilon_s$ breaks 
spin-rotation invariance, so by increasing $\epsilon_s$ we cross over 
to class $D$III.  By increasing both $\epsilon_s$ and $\epsilon_t$ we
break all symmetries and cross over to class $D$.  We call a symmetry
``maximally broken'' when its symmetry-breaking parameter
($\epsilon_s$ or $\epsilon_t$) equals unity.  Whenever a symmetry is
either unbroken or maximally broken, the probability distribution of
the Gaussian ensemble can be presented in the simple form
(\ref{gaussian}), with the corresponding symmetry constraints imposed
on ${\cal H}$.

All information about the level statistics is contained in the joint
probability distribution for the eigenvalues, $P(\{\omega\})$.  This
distribution is a complicated function in general, but it takes a
simple form for each universality class. By diagonalizing the
BdG-Hamiltonian
      \[
      {\cal H} = U \pmatrix{\omega &0\cr 0 &-\omega\cr} U^{-1},
      \quad \omega = {\rm diag}(\omega_1,\omega_2,...)
      \]
and computing the Jacobian of the transformation to diagonal form, we
obtain the formula
      \begin{equation}
      P(\{\omega\})d\{\omega\}
      = \prod_{i < j} |\omega_i^2 - \omega_j^2|^\beta
      \prod_k |\omega_k|^\alpha e^{-\omega_k^2 / v^2}
      d\omega_k
      \label{jpdensity}
      \end{equation}
where, for the individual cases,
      \begin{eqnarray}
      &&{\rm class \ }D : \quad \beta = 2, \quad \alpha = 0;
      \nonumber \\
      &&{\rm class \ }C : \quad \beta = 2, \quad \alpha = 2;
      \nonumber \\
      &&{\rm class \ }D{\rm III} : \quad \beta = 4, \quad \alpha = 1;
      \nonumber \\
      &&{\rm class \ }C{\rm I} : \quad \beta = 1, \quad \alpha = 1.
      \nonumber
      \end{eqnarray} 
These expressions for $P(\{\omega\})$ can be derived by 
elementary means. A particularly elegant derivation uses the 
interpretation of ${\cal H}$ as being tangent to the symmetric space 
of type $D$, $D$III, $C$, or $C$I.  The Jacobian can then be read off
immediately from the tabulated root systems of these spaces.

The formula for $P(\{\omega\})$ permits some immediate conclusions to
be drawn. Clearly, the significance of the parameter $\alpha$ is that
it governs the level repulsion from the origin $\omega = 0$, while
$\beta$ gives the mutual repulsion between levels.  For the following
it is useful to view the factor $|\omega_k|^\alpha$ as being due to
the interaction of the $k^{\rm th}$ level with its ``image'' at
$-\omega_k$.  Similarly we view the factor $\omega_i+\omega_j$ in
$\omega_i^2-\omega_j^2$ as resulting from the interaction of the
$i^{\rm th}$ level with the image of the $j^{\rm th}$ level.  At
energies $\omega$ much greater than the mean level spacing, the
interaction of levels with their distant images at negative energies
is expected to be irrelevant.  Therefore the level statistics derived
from (\ref{jpdensity}) will reduce, in that limit, to the usual
Wigner-Dyson statistics as determined by the parameter $\beta$.  On
the other hand, in the opposite limit of energies of order unity on
the scale set by the level spacing, the level statistics will be
different from Wigner-Dyson.  In particular, by the definition of
$P(\{\omega\})$ as a joint probability density we immediately conclude
that the mean density of levels near zero behaves as
      \begin{equation}
      \langle\rho(\omega)\rangle = 
      \langle {\rm Tr}\delta(\omega-{\cal H}) \rangle
      \sim |\omega|^\alpha \quad (\omega\to 0).
      \label{rho}
      \end{equation}
Note that for the systems where our random-matrix description applies,
the exponent $\alpha$ is predicted to be universal, dependent only on
symmetry.  The value of $\alpha$ for the symmetry classes $C$I and $C$
is easily understood from the fact that the repulsion of a level from
its own image is caused by the pairing field $\Delta$. For class $C$
pairing matrix elements are complex, whereas for class $C$I all
pairing matrix elements can be chosen to be real.  By a standard power
counting argument this results in $\alpha = 2$ and $\alpha = 1$
respectively.  To understand why $\alpha$ is zero for class $D$, note
that in this case a level and its own image are not really physically
distinct but are copies of the {\it same} single-electron state.  (In
contrast, for the classes $C$ and $C$I the hole level has its spin
flipped relative to the particle level.)  The pairing matrix element
between identical states vanishes by the Pauli principle -- or put
differently, the matrix $\Delta$ in (\ref{BdG}) is antisymmetric and
therefore has zeroes on its diagonal --, which results in $\alpha =
0$.

\section{Spectral Statistics}
\label{sec:spectrum}

\subsection{Exact Results}
\label{sec:spectral_exact}

Our interest here is in the level correlations for a large matrix
dimension $2N$.  These are easy to compute when the symmetry class is
$C$ or $D$.  Consider first class $C$.  For this symmetry class we can
interpret $P(\{\omega\})$ as the joint probability density of a
Gaussian Unitary Ensemble (GUE) of $2N$ levels $\omega_1, \omega_{\bar
1}, ..., \omega_N, \omega_{\bar N}$ subject to the mirror constraint
$\omega_{\bar k} = - \omega_{k}$.  The GUE joint probability density,
in turn, can be interpreted as the square of the ground state
wavefunction for a system of spinless nonrelativistic noninteracting
1D fermions confined by a harmonic well\cite{sutherland}.  This
correspondence of levels and Fermi particles turns the $n$-level
correlation functions of the GUE into the $n$-point static density
correlation functions of the Fermi system.  In the large-$N$ limit,
the spatial variation of the harmonic confining potential becomes
(locally) negligible and the gas of fermions can be treated as locally
free.  The mirror constraint means that whenever a fermion approaches
zero, then so does its mirror image.  Because the Pauli principle
makes the wavefunction vanish as two fermions approach each other,
this amounts to hard wall (or Dirichlet) boundary conditions at
$\omega = 0$. Hence we can compute the level density and its
correlations as the particle density and its correlations for a free
1D Fermi gas with Dirichlet boundary conditions at the origin.  The
free-fermion wavefunctions that vanish at $\omega = 0$ are
$\sin(\omega\tau)$, where $\tau$ plays the role of a ``wavenumber''.
By summing over the Fermi sea of states occupied in the ground state,
we obtain for the mean density of levels 
        \begin{equation} 
        \langle\rho(\omega)\rangle = {2\over\pi} \int_0^{\pi} 
        \sin^2(\omega\tau) d\tau = 
        1 - {\sin 2\pi\omega \over 2\pi\omega} .  
        \label{DOSforC}
        \end{equation} 
Here and throughout this subsection we follow the
convention of measuring $\omega$ in units of the mean spacing between
neighboring particles (i.e. of the level spacing) at a distance of
many spacings from zero.  Note that $\langle\rho(\omega)\rangle$ for
$\omega\to 0$ has the behavior expected from (\ref{rho}) (recall
$\alpha = 2$ for class $C$).  A similar calculation of the
density-density correlator of the Fermi gas yields the two-level
cluster function:
        \begin{eqnarray} &&\langle \rho(\omega_1)
        \rho(\omega_2) \rangle - \langle\rho(\omega_1)\rangle
        \langle\rho(\omega_2) \rangle \nonumber \\ &&- \big[
        \delta(\omega_1-\omega_2) + \delta(\omega_1 +\omega_2)\big]
        \langle\rho(\omega_1)\rangle \nonumber \\ = &&- \left[ {\sin
        \pi(\omega_1-\omega_2) \over \pi(\omega_1-\omega_2)} -
        {\sin\pi(\omega_1 +\omega_2) \over \pi(\omega_1+\omega_2)} 
        \right]^2.
        \nonumber 
        \end{eqnarray} 
Keeping $r = \omega_1-\omega_2 \not= 0$ fixed
and letting $\omega_1+\omega_2$ tend to infinity, we recover the
familiar GUE two-level cluster function $-\sin^2(\pi r)/(\pi r)^2$.
Similarly, all $n$-level functions $R_n(\omega_1, ...,
\omega_n) = \langle \rho(\omega_1)...\rho(\omega_n)\rangle$
can be calculated.  On subtracting the level self-correlations, 
which amounts to normal ordering in the particle-gas formulation, 
we obtain the result
       \begin{eqnarray}
       R_n(\omega_1,...,\omega_n) &=& {\rm Det} \left[ 
       \Psi_{C}(\omega_i,\omega_j) \right]_{i,j=1,...,n} \ ,
       \label{nlevels} \\
       \Psi_{C}(\omega_i,\omega_j) &=& {2\over\pi}
       \int_0^\pi \sin(\omega_i \tau) \sin(\omega_j \tau) d\tau ,
       \nonumber
       \end{eqnarray}
by simply using Wick's theorem for the free Fermi gas.

We turn to the symmetry class $D$. It is convenient again to use the
interpretation of the joint probability density as a Gaussian Unitary
Ensemble of $2N$ levels with a mirror constraint. The only change from
before is that the repulsion of a level from its own mirror image is
now absent ($\alpha = 0$).  Correspondingly the single-fermion
wavefunctions of the Fermi gas no longer vanish on approaching the
origin.  Instead, what we need to demand is that they be {\it even}
functions of $\omega$, which is the same as imposing vanishing
derivative (or Neumann) boundary conditions at $\omega = 0$. Thus the
level $\leftrightarrow$ particle correspondence now leads to the free
Fermi gas with Neumann boundary conditions at the origin.  Doing the
same kind of calculation as before we find
        \begin{equation}
        \langle \rho(\omega) \rangle = {2\over\pi}
        \int_0^{\pi} \cos^2(\omega\tau) d\tau = 
        1 + {\sin(2\pi\omega)\over 2\pi\omega} ,
        \label{DOSforD}
        \end{equation}
and the result (\ref{nlevels}) remains valid if we replace
$\Psi_{C}$ by $\Psi_{D}$,
       \[
       \Psi_{D}(\omega_i,\omega_j) = {2\over\pi} \int_0^\pi 
       \cos(\omega_i \tau) \cos(\omega_j \tau) d\tau .
       \]
{}From (\ref{DOSforD}) we see that for a metallic quantum dot with
spin-orbit scattering (class $D$), the proximity of a superconductor
with $\langle e^{i\phi} \rangle = 0$ {\it enhances} the level
density at the chemical potential!  While this effect may seem
physically surprising, it is very natural in the Fermi-gas picture of
the levels. The pressure of the gas pushes particles (or levels)
against the ``wall'' at $\omega = 0$.  Because it is the current
rather than the density that is required to vanish by the Neumann
boundary condition, an excess particle density forms at the wall such
that the extra statistical force balances the pressure. 
 
More effort is required by the symmetry classes $C$I and $D$III, where
$\beta = 1$ and $\beta = 4$.  It is still possible in these cases to
map the level statistics problem on a model of particles moving on a
half line, but progress is slowed down by the fact that the particles
now interact with each other.  By a standard transformation\cite{OP}
one can show that their motion is governed by the Hamiltonian of the
Calogero-Sutherland model (CSM) associated with the symmetric spaces
of type $C$I and $D$III.  For the case of the CSMs corresponding to
the Wigner-Dyson Ensembles, it was recently found\cite{hz} that the
CSM particles behave as a gas of {\it free anyons}, i.e. particles
with fractional charge and statistics.  Although we have some
preliminary results indicating that the free anyon gas picture can be
adapted to the present situation, the details have not been worked out
yet.

A quick way to get the infrared (or large-$\omega$) asymptotics of the
level density for $C$I and $D$III is to bosonize\cite{coleman} the
CSM.  This procedure has been argued\cite{kawakami} to yield the $c =
1$ conformal field theory of a free boson with compactification radius
$R = \sqrt{\beta/2}$.  The expression for the CSM particle density
$\bar\psi\psi$ in terms of the boson field $\varphi$
is\cite{mandelstam}
        \begin{equation}
        \bar{\psi}\psi = \partial_\omega \varphi + {\rm const} 
        \times \cos ( \sqrt{4\pi} \varphi /R  + k_F \omega ) .
        \label{bos_dens}
        \end{equation} 
(Recall that by the level $\leftrightarrow$ particle correspondence
$\omega$ is to be interpreted as a space coordinate here.)
The mirror constraint of the CSM for $C$I and $D$III translates 
into a boundary condition on $\varphi$ at $\omega = 0$.  Since the 
vertex operator $\exp(\sqrt{4\pi}i\varphi/R)$ has the scaling 
dimension $1/R^2 = 2/\beta$, we expect
       \begin{equation}
       \langle\rho(\omega)\rangle = 1 + \omega^{-2/\beta} 
       A_\beta(\omega) + ...
       \label{asymp_dens}
       \end{equation} 
where $A_\beta(\omega)$ is a function that oscillates with a period
determined by the mean spacing.  Note that (\ref{asymp_dens}) is
consistent with the $\beta = 2$ results (\ref{DOSforC}) and
(\ref{DOSforD}).  Note also that the first term on the right-hand side
of (\ref{bos_dens}) gives a vanishing contribution to the average
density, although it does contribute to the CSM density-density
correlator. This is because the current $\partial_\omega \varphi$, 
being linear in the boson field $\varphi$, has a vanishing expectation 
value even when there is a boundary. 

The validity of bosonization and conformal field theory arguments is
restricted to the infrared regime.  To obtain expressions that are 
valid in the {\it entire} range of frequencies $\omega$, we turn to
the orthogonal polynomial method of Dyson and Mehta\cite{mehta}.  
The substitution $x_k = \omega_k^2$ turns (\ref{jpdensity}) for 
$\alpha = 1$ into
     \[
      p(\{x\})d\{x\} = {\rm const} \times 
      \prod_{i < j} |x_i - x_j|^\beta
      \prod_k e^{-x_k / v^2} dx_k ,
      \]
which defines what has been called\cite{ns} the Laguerre Orthogonal
Ensemble (LOE) for $\beta = 1$, and the Laguerre Symplectic Ensemble
(LSE) for $\beta = 4$.  Note that this nomenclature is rather
unfortunate in the present context.  As we saw, the LOE relates to the
symmetric space ${\rm Sp}(2N)/{\rm U}(N)$, while the LSE relates to
the symmetric space ${\rm SO}(4N)/{\rm U}(2N)$.  In both cases the
invariance group is a {\it unitary} group, ${\rm U}(N)$ or ${\rm
  U}(2N)$.  Closed expressions for the $n$-level correlation functions
of these ensembles have recently been published by Nagao and
Slevin\cite{ns}. By using some identities for Bessel functions and
returning to the variable $\omega=x^2$, we can cast their results for
the mean density in the form 
       \begin{eqnarray}
       C{\rm I}: \quad \langle\rho(\omega)\rangle &=& 
       F(\pi\omega) , \nonumber \\
       D{\rm III}: \quad \langle\rho(\omega)\rangle &=& 
       F(2\pi\omega) + \pi J_1(2\pi\omega) / 2 , 
       \nonumber \\
       F(z) &=& {\pi\over 2} \int_0^z dt \ J_0(t)
       J_1(t) / t ,
       \nonumber
       \end{eqnarray}
where $J_k$ is the Bessel function of order $k$.  (Remember that we
are taking the level spacing at large $\omega$ for our energy unit.  
The levels are counted without multiplicity.)  From this we read off 
the small-$\omega$ expansions:
      \[
      \langle\rho(\omega)\rangle = \beta\pi^2 \omega/4
      + {\cal O}(\omega^3)  \qquad 
      (C{\rm I} \ {\rm and} \ D{\rm III}) .
      \]
Knowing the mean density, we can construct the full one-point function
$\langle g(\omega) \rangle = \langle {\rm Tr} (\omega + i\delta-{\cal
H})^{-1} \rangle$ by causality, i.e. by using the dispersion relation
that connects the real and imaginary parts of a holomorphic function 
on the upper complex half-plane.  The results can be presented in the 
form
      \begin{eqnarray}
      C{\rm I}: \quad &&\langle g(\omega) \rangle 
      \nonumber \\
      = &&-i\pi + i \int_1^\infty du \int_{-1}^{+1} dv \ 
      {\sqrt{1-v^2}\over \sqrt{u^2-1}} {e^{i\pi\omega(u-v)}
      \over u-v} ,
      \nonumber \\
      D{\rm III}: \quad &&\langle g(\omega) \rangle = 
      -i\pi + i\pi \int_1^\infty du \ {u\over\sqrt{u^2-1}}
      e^{2i\pi\omega u}
      \nonumber \\
      &&-i \int_1^\infty du \int_{-1}^{+1} dv \ 
      {\sqrt{u^2-1}\over \sqrt{1-v^2}} {e^{2i\pi\omega(u-v)}
      \over u-v} .
      \nonumber
      \end{eqnarray}
Although it is hard work to construct these expressions directly, 
they can easily be verified.  For that we simply differentiate the 
result for $\langle g(\omega) \rangle$ with respect to $\omega$, 
thereby cancelling the factor $u-v$ in the denominator of the double
integrals.  The integrals over $u$ and $v$ then separate, and on
taking the imaginary part each integral produces a Bessel function.  
By using standard recursion relations for these functions and 
then undoing the $\omega$-differentiation by integration, we
immediately retrieve the expressions for $\langle\rho(\omega)
\rangle$ given earlier.

The double integrals for $C$I ($\beta=1$) and $D$III ($\beta=4$) are
seen to be related by a duality transformation that exchanges the
compact ($v$) and noncompact ($u$) degrees of freedom.  A similar
duality relation holds for the conventional Wigner-Dyson ensembles
with $\beta = 1$ and $\beta = 4$\cite{efetov}.  In the limit of large
$\omega$ we get the following asymptotic expansions for the one-point 
function:
      \begin{eqnarray}
      C{\rm I}: \quad \langle g(\omega)\rangle &=&
      -i\pi - {1\over 2\omega} + {1\over 4\pi\omega^2}
      e^{2\pi i\omega} + ... \nonumber \\
      D{\rm III}: \quad \langle g(\omega)\rangle &=&
      -i\pi + {1\over 4\omega} + {i\pi\over 2\sqrt{\omega}}
      e^{2\pi i\omega + i\pi/4} + ...
      \nonumber
      \end{eqnarray}
For completeness, the one-point functions for the symmetry classes $C$
and $D$ ($\beta = 2$), as determined from (\ref{DOSforC}) and 
(\ref{DOSforD}) by causality, are
        \[
        \langle g(\omega) \rangle = - i\pi
        + (1-\alpha) {1 - e^{2\pi i\omega}\over 2\omega}
        \quad (C \ {\rm and} \ D).
        \]
By comparing with (\ref{asymp_dens}) we see that the oscillatory
correction to the stationary asymptotic limit $\langle g(\omega)
\rangle \to -i\pi$ agrees with what is expected from the conformal
limit of the Calogero-Sutherland model, in all cases. The smooth
($1/\omega$) part of the correction is purely real and does not enter
into the asymptotic expression for the density of states.

The authors of Ref.\cite{ns} subjected the $n$-level correlation
functions for $n > 1$ to a renormalization or unfolding procedure in
the low-frequency regime they call ``nonuniversal'' (meaning
different from standard Wigner-Dyson).  We wish to emphasize that such
a procedure is neither necessary nor appropriate here.  Both the mean
density and the level correlation functions {\it are universal as they
stand} -- the restriction to the class of system we have delineated
being understood, of course -- and are not to be corrupted by any kind
of unfolding.

\subsection{Diagrammatic Perturbation Theory for the One-Point 
Function}
\label{OnePoint}

As we have seen, $\langle g(\omega) \rangle$ tends to a constant for
frequencies much larger than the level spacing.  The leading {\it
  smooth} (i.e. nonoscillatory) correction is of order $1/\omega$ in
all cases.  More precisely, on restoring the physical units and taking
into account the multiplicity of levels, we have $\langle g(\omega)
\rangle = -i\pi\nu + c / \omega + {\cal O}(1/\omega^2)$, where $c =
-1$ for $C$ and $C$I, and $c = +1/2$ for $D$ and $D$III.  $\nu$ is the
asymptotic (i.e. large-$\omega$) density of states.  We are now going
to show how to obtain this result by a variant of the impurity diagram
technique, a method which has the attractive feature of generalizing
easily to the calculation of transport properties of an open system.
It also has the great virtue of lending itself to semiclassical
interpretation, which will help improve our understanding of the
physics involved.

\begin{figure}
\centerline{\psfig{figure=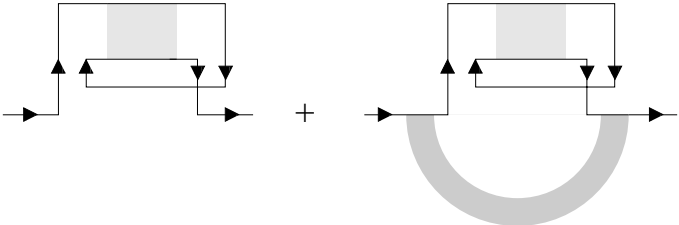,height=3.0cm}}
\caption{Diagrams contributing to the average single-particle
 Green's function.  The light-shaded regions represent a $D$-type
 cooperon mode, the dark-shaded one a nonsingular
 $\Pi_A^{{\rm d}0}$-ladder.}
\label{fig2}
\end{figure}

The impurity diagram technique in its present version starts from the
usual idea of expanding $(\omega+i\delta-{\cal H})^{-1}$ in a
geometric series with respect to ${\cal H}$ and then taking the
ensemble average.  Because ${\cal H}$ is Gaussian distributed, the
ensemble average is evaluated by forming all products of pairwise
contractions $\langle{\cal H}{\cal H}\rangle$, which are determined by
the basic law (\ref{correlator}).  To resum the relevant
contributions, we use standard diagrammatic techniques.  On
multiplying the factors on the right-hand side of (\ref{correlator})
we generate eight terms.  In explicit index notation these are given
by
       \begin{eqnarray}
       \left(\Pi_A^{{\rm d}0}\right)_{\alpha\gamma,\beta\delta}
       &=& \delta_{\alpha\delta} \delta_{\gamma\beta} ,
       \nonumber \\
       \left(\Pi_A^{{\rm d}1}\right)_{\alpha\gamma,\beta\delta}
       &=& \sum_k (J_k)_{\alpha\delta} (J_k)_{\gamma\beta} ,
       \nonumber \\
       \left(\Pi_A^{{\rm c}0}\right)_{\alpha\gamma,\beta\delta}
       &=& \tau_{\alpha\gamma} (\tau^{-1})_{\delta\beta} ,
       \nonumber \\
       \left(\Pi_A^{{\rm c}1}\right)_{\alpha\gamma,\beta\delta}
       &=& \sum_k (J_k \tau)_{\alpha\gamma} 
       (\tau^{-1}J_k)_{\delta\beta} ,
       \nonumber \\
       \left(\Pi_D^{{\rm c}0}\right)_{\alpha\gamma,\beta\delta}
       &=& - (\Sigma_x)_{\gamma\alpha} (\Sigma_x)_{\beta\delta} ,
       \label{contractions} \\
       \left(\Pi_D^{{\rm c}1}\right)_{\alpha\gamma,\beta\delta}
       &=& - \sum_k (J_k\Sigma_x)_{\gamma\alpha} 
       (\Sigma_x J_k)_{\beta\delta} ,
       \nonumber \\
       \left(\Pi_D^{{\rm d}0}\right)_{\alpha\gamma,\beta\delta}
       &=& - (\Sigma_x \tau)_{\beta\gamma} 
       (\tau^{-1} \Sigma_x)_{\delta\alpha} ,
       \nonumber \\
       \left(\Pi_D^{{\rm d}1}\right)_{\alpha\gamma,\beta\delta}
       &=& - \sum_k (\Sigma_x J_k \tau)_{\beta\gamma} 
       (\tau^{-1} J_k \Sigma_x)_{\delta\alpha} .
       \nonumber
       \end{eqnarray} 
It is characteristic of the contractions indexed by the letter ``c''
that the initial states $\beta, \delta$ bear a definite relation to
each other, and so do the final states $\alpha, \gamma$.  This
situation is reminiscent of the cooperon channel of disordered
mesoscopic systems where a pair of particles with initial momenta $p$
and $-p$ are scattered to final momenta $p'$ and $-p'$.  Similarly,
the contractions indexed by ``d'' correspond to the diffuson channel
where a pair with momenta $p,p'$ is scattered to a pair with momenta
$p',p$.  The contractions with subscript ``$D$'' owe their existence
to the operation of particle-hole conjugation $X \mapsto -\Sigma_x
X^{\rm T} \Sigma_x$, whose fixed point set is the orthogonal algebra
$D_{2N} \equiv {\rm so}(4N)$.  The name of the ``$A$''-type
contractions is motivated by the fact that they determine the
second moments of the conventional Wigner-Dyson ensembles describing
N-systems (without any coupling of particles and holes), which derive
from the unitary algebra $A_{k-1} \equiv {\rm su}(k)$.  The numerals 0 
and 1 distinguish between spin-singlet and spin-triplet contractions.
Using the conventions (\ref{contractions}) we can write the 
correlation law (\ref{correlator}) in the form
      \begin{eqnarray}
      \langle {\cal H}_{\alpha\beta} {\cal H}_{\gamma\delta}
      \rangle &&= v^2 \Pi_{\alpha\gamma,\beta\delta} ,
      \nonumber \\
      \Pi &&= \Pi_A^{{\rm d}0} + \Pi_D^{{\rm c}0} 
      \nonumber \\
      &&+ (1-\epsilon_t) (\Pi_A^{{\rm c}0} + \Pi_D^{{\rm d}0})
      \label{decompose} \\
      &&+ (1-\epsilon_s)(\Pi_A^{{\rm d}1}+\Pi_D^{{\rm c}1})
      \nonumber \\
      &&+ (1-\epsilon_s)(1-\epsilon_t)(\Pi_A^{{\rm c}1} + 
      \Pi_D^{{\rm d}1}) .
      \nonumber
      \end{eqnarray}
Our goal is to find the large-$\omega$ behavior of $\langle
g(\omega) \rangle$.  What are the dominant diagrams in this limit?
{}From what has been said, the $A$-type contractions give rise to
Wigner-Dyson statistics, whereas the $D$-type contractions are
responsible for the corrections to it.  Since our systems are
Wigner-Dyson in the limit of large $\omega$, the $D$-type contractions 
must become irrelevant in that limit.  Moreover, in the Wigner-Dyson
regime the average Green's function is known to be featureless and
independent of the symmetry-breaking perturbations $\epsilon_s$ and
$\epsilon_t$.  We therefore conclude that $\langle g(\omega)
\rangle$ is completely determined by $\Pi_A^{{\rm d}0}$-contractions
for $\omega\to\infty$ (and large $N$).  By summing all nested
$\Pi_A^{{\rm d}0}$ self-energy graphs, we get Pastur's approximation
$G^0$ to $G \equiv \langle (\omega+i\delta-{\cal H})^{-1} \rangle$:
        \begin{equation}
        G^0 = (\omega+i\delta - v^2 {\rm Tr}G^0)^{-1} .
        \label{pastur}
        \end{equation}
This equation is exact for $N \to \infty$ and large $\omega$.  Its 
solution yields Wigner's semicircle law for the density of states. 
Putting $v^2 = \lambda^2 / 4N$ and focusing on the central region of 
the semicircle, we obtain
        \[
        {\rm Tr} G^0 = -i\pi\nu + (\pi\nu)^2 \omega / 8N +
        {\cal O}(\omega^2 / N^2)
        \]
where $\nu = 4N/\pi\lambda$ is identified as the asymptotic density of
states.  What we need to do to probe the local structure of the
spectrum, is to keep the product $\nu\omega$ fixed while sending $N$
to infinity.  The corrections to ${\rm Tr}G^0 = -i\pi\nu$ from
Pastur's equation are seen to become negligible in this limit.
However, we know that corrections to the stationary asymptotic
behavior ${\rm Tr}G^0 = -i\pi\nu$ do appear as we approach zero
frequency.  These must be due to the contractions of type $D$. The
leading correction is depicted in Fig.~\ref{fig2}, where the 
light-shaded regions represent ladder diagrams built either from
$\Pi_D^{{\rm c}0}$-contractions or from $\Pi_D^{{\rm
c}1}$-contractions. 
\narrowtext
The sum of the former diagrams, which we call the $D$-type 
spin-singlet cooperon mode and denote by $S_{\alpha\gamma,
\beta\delta}^0$, has the expression
        \[
        S^0 = v^2 \Pi_0/ (1 - v^2 K \Pi_0)
        \]
with $K_{\alpha\gamma,\beta\delta} = \delta_{\alpha\beta} 
\delta_{\gamma\delta} G_{\alpha\alpha}^0 G_{\gamma\gamma}^0$
and $\Pi_0 \equiv \Pi_D^{{\rm c}0}$.  Its large-$N$ limit is
        \begin{equation}
        S^0 = {\lambda^2 \Pi_0 \over 
        - i\pi\nu\omega} + {\cal O}(1/N).
        \label{ladder_Dc0}
        \end{equation}
Similarly, the sum of all ladder diagrams built from $\Pi_D^{{\rm c} 
1}$-contractions, the $D$-type spin-triplet cooperon, is evaluated as
        \begin{eqnarray}
        S^1 &=& (1-\epsilon_s) v^2 \Pi_1 / 
        [1 - (1-\epsilon_s) v^2 K \Pi_1 ]
        \nonumber \\
        &=& {\lambda^2 \Pi_1 \over
        \eta_s - i\pi\nu\omega} + {\cal O}(1/N)
        \label{ladder_Dc1}
        \end{eqnarray}
where $\eta_s = 4N\epsilon_s$ and $\Pi_1 \equiv \Pi_D^{{\rm c}1}$.
The dependence of $S^1$ on the parameter $\epsilon_s$ through the
product $N\epsilon_s$ means that the breaking of spin-rotation
invariance takes place on scales $\epsilon_s \sim 1/N$ and thus is
very fast.  This rapid crossover happens because the crossover scale
is determined by the typical size of a symmetry-breaking matrix
element in relation to the level spacing, which is $\nu^{-1} =
\lambda\pi/4N$ for our choice of normalization.

Note that the expressions (\ref{ladder_Dc0}) and (\ref{ladder_Dc1})
are singular at $\omega = 0 = \eta_s$ even though the sums of ladder
diagrams they represent are built from retarded Green's functions only
($G^+ G^+$ channel).  This is a novel feature which does not occur for
the standard Wigner-Dyson ensembles, where singular ladders exist only
in the advanced-retarded (or $G^- G^+$) channel.  The singularity in
the present case comes about because the minus sign from
$K_{\alpha\gamma, \alpha\gamma} = - 1/\lambda^2$ is cancelled by a
minus sign appearing in the definition of the contractions of type
$D$, thereby turning an alternating (conditionally convergent) series
into a divergent one.

The dark-shaded region appearing in the second diagram of
Fig.~\ref{fig2} represents a $\Pi_A^{{\rm d}0}$-ladder. According to
(\ref{contractions}), the contractions of type $A$ come with an
overall plus sign, so the minus signs now do {\it not} cancel, and the
ladder sum is always finite.  Computing the sum we find that this
nonsingular $\Pi_A^{{\rm d}0}$-ladder renormalizes the first diagram
in Fig.~\ref{fig2} by a factor of $1-(1-1+1-...) = 1/(1+1) = 1/2$.
(We mention in passing that nonsingular ladders of this kind are the
random-matrix analog of the single-impurity lines that appear in the
context of the impurity diagram technique.)

To evaluate the first diagram of Fig.~\ref{fig2} we need the
following sums:
        \begin{eqnarray}
        &&\sum_\beta \left( \Pi_D^{{\rm c}0} \right)_{\alpha\beta,
        \beta\alpha} = - \sum_\beta 
        \left( \Sigma_x \right)_{\beta\alpha}
        \left( \Sigma_x \right)_{\beta\alpha} = - 1 ,
        \nonumber \\
        &&\sum_\beta \left( \Pi_D^{{\rm c}1} \right)_{\alpha\beta,
        \beta\alpha} = - \sum_\beta 
        \left( \Sigma_x J_k \right)_{\beta\alpha}
        \left( J_k \Sigma_x \right)_{\beta\alpha} = + 3 .
       \nonumber
       \end{eqnarray}
We then obtain
        \[
        \langle g(\omega) \rangle = -i\pi\nu
        +i\pi\nu \left( {3/2 \over \eta_s - \pi i\nu\omega}
        - {1/2 \over -\pi i\nu\omega} \right) + ... ,
        \]
which agrees with the result stated at the beginning of the current
subsection for $\eta_s = 0$ (classes $C$ and $C$I) and
$\eta_s \to \infty$ (classes $D$ and $D$III).  For the classes
$C$ and $D$, smooth corrections of higher order in $1/\omega$ are
completely absent from the exact result of Sec.~\ref
{sec:spectral_exact}.  This implies that all diagrams of higher order
than the ones considered here must cancel each other, in these two
cases.  (The oscillatory correction $\sim A_\beta \omega^{-2/\beta}$
is nonanalytic in the expansion parameter $1/\omega$ and therefore
remains invisible to all orders of perturbation theory.)

\section{Slow Modes}
\label{sec:modes}

In all previous Green's function treatments of NS-systems the diagrams
were enumerated by the number of Andreev reflections.  Unfortunately,
when the perturbation expansion is organized in that way, the vast
number of possibilities to insert Andreev reflections into the
diagrams generates a flood of terms which is hard to control, and as a
result it is very easy to miss important contributions.  The technical
innovation made in the present paper is not to single out Andreev
reflections but treat them on exactly the same footing as the
processes of impurity scattering.  This is possible by our dynamical
assumptions ensuring that the quantum mechanical phase acquired during
Andreev reflection, can be regarded as a random variable with zero
mean.  Our key technical step is the decomposition (\ref{decompose})
which leads to an organization of the perturbation-theory diagrams by
{\it symmetry}.  In the preceding subsection we discussed how the
contractions $\Pi_D^{{\rm c}0}$ and $\Pi_D^{{\rm c}1}$ generate
singular geometric series of ladder diagrams.  In the same way, every
one of the other contractions gives rise to one singular ladder.
These singular modes can be visualized as follows:
\begin{eqnarray}
\unitlength0.04cm
&
\begin{minipage}{7.0cm}
\begin{picture}(200,70)(0,0)
\put(20,20){\vector(1,0){10}}
\put(30,20){\vector(1,0){20}}
\put(50,20){\vector(1,0){20}}
\put(70,20){\line(1,0){10}}
\put(20,40){\vector(1,0){10}}
\put(30,40){\vector(1,0){20}}
\put(50,40){\vector(1,0){20}}
\put(70,40){\line(1,0){10}}
\multiput(40,20)(0,2){10}{\line(0,1){1}}
\multiput(60,20)(0,2){10}{\line(0,1){1}}
\put(120,20){\vector(1,0){10}}
\put(130,20){\vector(1,0){20}}
\put(150,20){\vector(1,0){20}}
\put(170,20){\line(1,0){10}}
\put(120,40){\vector(1,0){10}}
\put(130,40){\vector(1,0){20}}
\put(150,40){\vector(1,0){20}}
\put(170,40){\line(1,0){10}}
\multiput(140,20)(0,2){10}{\line(0,1){1}}
\multiput(160,20)(0,2){10}{\line(0,1){1}}
\put(05,20){\makebox(0,0){${\rm p(h)}$}}
\put(05,40){\makebox(0,0){${\rm p(h)}$}}
\put(100,20){\makebox(0,0){$\dots$}}
\put(100,40){\makebox(0,0){$\dots$}}
\put(100,50){\makebox(0,0){$G^+(G^-)$}}
\put(100,10){\makebox(0,0){$G^-(G^+)$}}
\end{picture}
\end{minipage}&\nonumber\\
&A{\rm -type}\;{\rm cooperon,}&\nonumber\\
%
%
&
\begin{minipage}{7.0cm}
\begin{picture}(200,70)(0,0)
\put(80,20){\vector(-1,0){10}}
\put(70,20){\vector(-1,0){20}}
\put(50,20){\vector(-1,0){20}}
\put(30,20){\line(-1,0){10}}
\put(20,40){\vector(1,0){10}}
\put(30,40){\vector(1,0){20}}
\put(50,40){\vector(1,0){20}}
\put(70,40){\line(1,0){10}}
\multiput(40,20)(0,2){10}{\line(0,1){1}}
\multiput(60,20)(0,2){10}{\line(0,1){1}}
\put(180,20){\vector(-1,0){10}}
\put(170,20){\vector(-1,0){20}}
\put(150,20){\vector(-1,0){20}}
\put(130,20){\line(-1,0){10}}
\put(120,40){\vector(1,0){10}}
\put(130,40){\vector(1,0){20}}
\put(150,40){\vector(1,0){20}}
\put(170,40){\line(1,0){10}}
\multiput(140,20)(0,2){10}{\line(0,1){1}}
\multiput(160,20)(0,2){10}{\line(0,1){1}}
\put(05,20){\makebox(0,0){${\rm p(h)}$}}
\put(05,40){\makebox(0,0){${\rm p(h)}$}}
\put(100,20){\makebox(0,0){$\dots$}}
\put(100,40){\makebox(0,0){$\dots$}}
\put(100,50){\makebox(0,0){$G^+(G^-)$}}
\put(100,10){\makebox(0,0){$G^-(G^+)$}}
\end{picture}
\end{minipage}&\nonumber\\
&A{\rm -type}\;{\rm diffuson,}&\nonumber\\
%
%
&
\begin{minipage}{7.0cm}
\begin{picture}(200,70)(0,0)
\put(20,20){\vector(1,0){10}}
\put(30,20){\vector(1,0){20}}
\put(50,20){\vector(1,0){20}}
\put(70,20){\line(1,0){10}}
\put(20,40){\vector(1,0){10}}
\put(30,40){\vector(1,0){20}}
\put(50,40){\vector(1,0){20}}
\put(70,40){\line(1,0){10}}
\multiput(40,20)(0,2){10}{\line(0,1){1}}
\multiput(60,20)(0,2){10}{\line(0,1){1}}
\put(120,20){\vector(1,0){10}}
\put(130,20){\vector(1,0){20}}
\put(150,20){\vector(1,0){20}}
\put(170,20){\line(1,0){10}}
\put(120,40){\vector(1,0){10}}
\put(130,40){\vector(1,0){20}}
\put(150,40){\vector(1,0){20}}
\put(170,40){\line(1,0){10}}
\multiput(140,20)(0,2){10}{\line(0,1){1}}
\multiput(160,20)(0,2){10}{\line(0,1){1}}
\put(05,20){\makebox(0,0){${\rm h(p)}$}}
\put(05,40){\makebox(0,0){${\rm p(h)}$}}
\put(100,20){\makebox(0,0){$\dots$}}
\put(100,40){\makebox(0,0){$\dots$}}
\put(100,50){\makebox(0,0){$G^+(G^-)$}}
\put(100,10){\makebox(0,0){$G^+(G^-)$}}
\end{picture}
\end{minipage}&\nonumber\\
&D{\rm -type}\;{\rm cooperon,}&\nonumber\\
%
%
&
\begin{minipage}{7.0cm}
\begin{picture}(200,70)(0,0)
\put(80,20){\vector(-1,0){10}}
\put(70,20){\vector(-1,0){20}}
\put(50,20){\vector(-1,0){20}}
\put(30,20){\line(-1,0){10}}
\put(20,40){\vector(1,0){10}}
\put(30,40){\vector(1,0){20}}
\put(50,40){\vector(1,0){20}}
\put(70,40){\line(1,0){10}}
\multiput(40,20)(0,2){10}{\line(0,1){1}}
\multiput(60,20)(0,2){10}{\line(0,1){1}}
\put(180,20){\vector(-1,0){10}}
\put(170,20){\vector(-1,0){20}}
\put(150,20){\vector(-1,0){20}}
\put(130,20){\line(-1,0){10}}
\put(120,40){\vector(1,0){10}}
\put(130,40){\vector(1,0){20}}
\put(150,40){\vector(1,0){20}}
\put(170,40){\line(1,0){10}}
\multiput(140,20)(0,2){10}{\line(0,1){1}}
\multiput(160,20)(0,2){10}{\line(0,1){1}}
\put(05,20){\makebox(0,0){${\rm h(p)}$}}
\put(05,40){\makebox(0,0){${\rm p(h)}$}}
\put(100,20){\makebox(0,0){$\dots$}}
\put(100,40){\makebox(0,0){$\dots$}}
\put(100,50){\makebox(0,0){$G^+(G^-)$}}
\put(100,10){\makebox(0,0){$G^+(G^-)$}}
\end{picture}
\end{minipage}&\nonumber\\
&D{\rm -type}\;{\rm diffuson.}&\nonumber
\end{eqnarray}
The dotted vertical lines represent both impurity scatterings and 
Andreev reflections, and they denote any one of the eight contractions
$\Pi_X^{\rm xS}$ ($X = A, D$; ${\rm x} = {\rm c,d}$; ${\rm S} = 0,1$).
The type of contraction is invariant within one ladder.  The $A$-type
modes are built from states of identical charge (two BdG-particles or 
two BdG-holes) propagating on opposite segments of the ladder, whereas 
the $D$-type modes are built from charge-reversed states (one particle 
and one hole).  The former are singular in the $G^+ G^-$ channel, the
latter in the $G^+ G^+$ (or $G^- G^-$) channel.  The arrows on the
Green's function lines indicate the order in which single-particle
states are visited.  For the cooperon modes the order on both lines is
the same, while for the diffuson modes it is reversed.  In the limit
$\omega = \eta_s = \eta_t = 0$ (with $\eta_t = 4N\epsilon_t$) all
modes are singular, or massless.  The $D$-type modes are made massive
by frequency (or voltage) $\omega$ while the $A$-type modes are
insensitive to such a perturbation.  The $A$-type cooperon and the
$D$-type diffuson are made massive by the breaking of time-reversal
symmetry.  Since a Green's function line carries spin 1/2, the modes
decompose into spin-singlet and spin-triplet ones.  The spin-triplet
modes are sensitive to spin-orbit scattering while the spin-singlet
modes are not.

We wish to mention that there is some redundancy in our classification
of modes, as the basic particle-hole symmetry (\ref{orthogonal}) 
causes the existence of certain relations among the matrix elements
of the Gorkov Green's function $G^{\pm}(\omega) = (\omega \pm 
i\varepsilon - {\cal H})^{-1}$.  In particular, the particle-particle 
and hole-hole matrix elements are related by 
        \begin{equation}
        G_{\rm pp}^{\pm}(\omega) = 
        - G_{\rm hh}^\mp(-\omega)^{\rm T} .
        \label{ph-relate}
        \end{equation}  
Similarly, $G_{\rm ph}^\pm(\omega) = - G_{\rm hp}^\mp(-\omega)^{\rm
T}$.  These identities transcribe into relations connecting the
singular modes.  For example, by using (\ref{ph-relate}) on one of the
Green's function lines of the $D$-type cooperon, we can make this mode
look like the $A$-type diffuson, at the expense of having to change
the sign of one frequency $(\omega\to -\omega)$.  In a similar way,
the $D$-type diffuson is related to the $A$-type cooperon (again with
a sign change in one of the frequencies).  In spite of that, we prefer
to treat the $A$- and $D$-type modes as separate entities.  The main
reason for doing so is that they respond differently to translations
of the energy: while the $D$-type modes are made massive by shifting 
the energy, the $A$-type modes are not.

In the present paper we restrict our considerations to the
ergodic (or zero-dimensional) limit.  To go beyond, we should
associate with each $A$-type and $D$-type mode a small momentum
variable (``slow modes'') and sum over momenta.  In this way,
it will not be difficult to generalize our results beyond the
ergodic limit.

\section{Weak Localization}
\label{sec:wl}

\begin{figure}
        \centerline{\psfig{figure=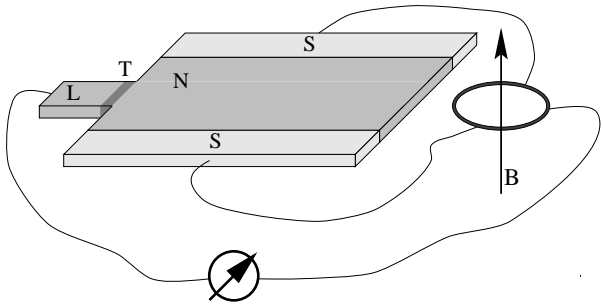,height=4.0cm}}
        \vspace{0.5cm}
        \caption{Andreev quantum dot (N) coupled to a single lead (L) 
                via a tunnelling barrier (T).  The flux loop on the 
                right is introduced to adjust the difference of the 
                order parameter phases of the superconducting regions 
                (S) to the value $\phi_1-\phi_2 = \pi$ 
                (cf. the discussion of Sec.~\ref{sec:dyn_input}).}
        \label{VIIfig1a}
        \end{figure}

Having made a thorough analysis of the isolated Andreev quantum dot,
we now turn to the discussion of the associated {\it open} system and
its transport properties.  To open up the dot in the simplest possible
way, we couple it to a single lead with $2M$ open channels (the factor
of 2 accounts for the spin degree of freedom), see Fig.~\ref{VIIfig1a}.
The transmission of charge excitations from the lead to the interior
of the dot is modelled by a set of (spin-independent) real hopping
matrix elements $W_{\mu a}$, where the index $a=1,\dots,M$
($\mu=1,\dots,N$) enumerates the channels carried by the lead (the
sites of the dot).  We assume $N \gg M \gg 1$.  Let $g$ denote the
conductance measured in units of $e^2 / h$.  To calculate $g$, we
employ the Landauer-B\"uttiker formula as generalized to NS-systems by
Takane and Ebisawa\cite{takane92},
        \begin{equation}
        g = 2
        \sum_{\tilde{a},\tilde{b}}|S_{\tilde{b}\tilde{a}
        }^{\rm hp}|^2,
        \label{VIIlandauer}
        \end{equation}
where the composite label $\tilde{a}=(a,s_a)$ comprises the spin
$s_a=\pm1/2$ and the index $a$ of an open channel in the lead, and
$S_{\tilde{b}\tilde{a}}^{\rm hp}$ denotes the scattering amplitude
connecting a particle coming in in channel $\tilde{a}$ with a hole
going out in channel $\tilde{b}$.  The $S$-matrix is given by
        \begin{equation}
        S_{\tilde{b}\tilde{a}}^{\rm hp} = 
        -2i W^{\rm T}_{\;\;b \mu } 
        G_{(\mu,s_b,{\rm h}),(\nu,s_a,{\rm p})} W_{\nu a},
        \label{VIISmat}
        \end{equation}
where 
        \[
        G=(i\delta-{\cal H} + i WW^{\rm T})^{-1}
        \]
is the Gorkov Green's function evaluated at the chemical potential. 
Without loss of generality, we may assume the matrices $W=\{W_{\mu 
a}\}$ to be of the form
        \begin{equation}
        W_{\mu a}=\gamma^{1/2}\delta_{\mu a} \ (\mu=1,\dots,N;
        a=1,\dots,M).
        \label{VIIWdiagonal}
        \end{equation}
The unitary transformation necessary to transform $W$ to the form
(\ref{VIIWdiagonal}) can be absorbed in the Hamiltonian
$\cal{H}$ by the invariance properties of the random-matrix ensemble
(\ref{correlator}).  Combining Eqs. (\ref{VIISmat}) and
(\ref{VIIlandauer}) and making use of (\ref{VIIWdiagonal}) we obtain
        \begin{equation}
        g = 8 \sum_{\mu s \mu' s'}
        \Gamma_{\mu} \left| G_{(\mu,s,{\rm h}),
        (\mu',s',{\rm p})} \right|^2 \Gamma_{\mu'},
        \label{VIIggreen}
        \end{equation}
where
        \[
        \Gamma_{\mu}=\left\{\begin{array}{ll}
        \gamma& \ {\rm for} \ \mu\le M,\\
        0& \ {\rm else.}
        \end{array}\right.
        \]
We are going to calculate this expression to leading order in the
small parameters $1/N$, $M/N$ and next-to-leading order in
$1/M$\cite{footnote1}.  Owing to the presence of the BdG particle-hole
degree of freedom, an analysis of Eq.~(\ref{VIIggreen}) within the
framework of plain diagrammatic perturbation theory turns out to
involve a sizable number of diagrams.  It is more efficient to
pre-analyze (\ref{VIIggreen}) by means of a set of exact identities
(Ward identities) before turning to diagrammatic methods.  This
calculation is detailed in Appendix \ref{app1}.  Here we restrict
ourselves to a presentation of the results and their interpretation in
terms of semiclassical trajectories.

Schematically, the conductance can be represented as:
        \begin{equation}
        g=
        \begin{minipage}{7.0cm}
        \psfig{figure=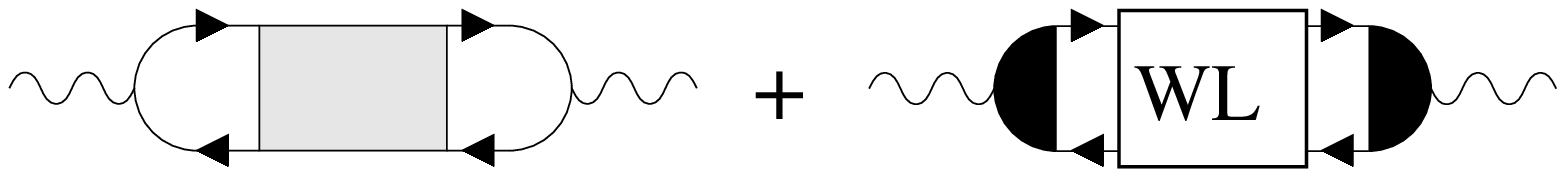,height=0.8cm}
        \end{minipage}
        \label{VIIgdiag}
        \end{equation}
where the wavy lines stand for the quantities $\{\Gamma_{\mu}\}$, the
shaded region denotes the singular $A$-type diffuson mode introduced
in Sec.~\ref{sec:modes},
        \[
        \psfig{figure=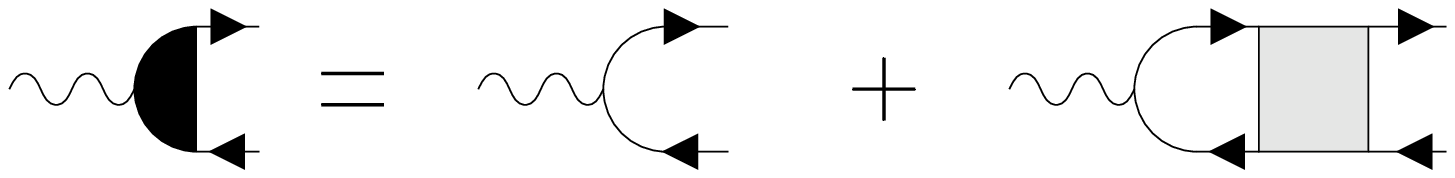,height=0.6cm} ,
        \]
and a summation over indices is understood.  The WL-building block
represents a quantum interference correction (the NS-analog of the
well-known weak localization correction for normal metals) to the
classical conductance.  In contrast with the pure N-case, however,
{\it two} qualitatively different processes contribute to the weak
localization correction for the Andreev dot:
        \[
        \centerline{\psfig{figure=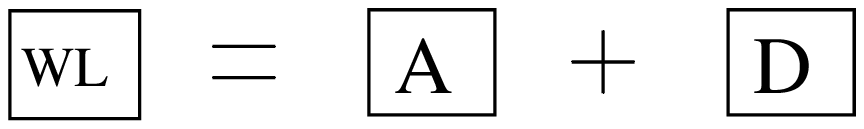,height=0.7cm}.}
        \]
Here the $A$($D$)-block is due to the presence of singular
$A$($D$)-type modes. Whereas the $A$-type contribution resembles the
standard weak localization correction known from normal metals, the
$D$-term does not have any analog in pure N-systems and is of a
different nature.  In the following we discuss separately the
classical conductance (the first diagram in (\ref{VIIgdiag})), the
$A$-type correction, and the $D$-type correction.

{\it Classical conductance:} Qualitatively speaking, the conductance 
is given by
        \begin{equation}
        g = \left|\sum_i A_i\right|^2,
        \label{VIIampprod}
        \end{equation}
where $A_i$ is the amplitude to traverse a certain scattering sequence
(indexed by $i$) connecting an incoming particle channel with an
outgoing hole channel.  The classical value of the conductance, $g_0$,
is obtained by evaluating the {\it incoherent} sum
        \[
        g_0=\sum_i\left| A_i\right|^2.
        \]
Quantitatively, we obtain 
        \[
        g_0=2MT,
        \]
where the transmission coefficient $T$ is the probability for an
electron incident from the lead to enter the dot instead of being
reflected back into the lead\cite{footnote2}.  This result is easy to
understand.  By the ergodicity of our system, an electron leaves the
dot with equal ($1/2$) probability as a particle or as a hole.  In the
latter case, two elementary charges are transferred across the entire
system.  Thus the dimensionless conductance per channel is $2 \times
1/2 \times T = T$.  Multiplying by the number of channels we get $g_0
= 2MT$.

\begin{figure}
        \centerline{\psfig{figure=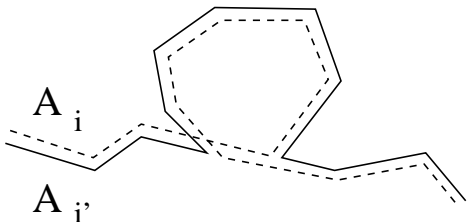,height=3.0cm}}
        \caption{Pair of semiclassical paths contributing to the 
        $A$-type weak localization process.}
        \label{VIIfig1}
        \end{figure}

{\it $A$-type corrections:} Weak localization corrections to the
classical conductance originate from the phase-coherent contributions
of nonidentical paths to the sum of amplitudes (\ref{VIIampprod}).  In
the case of the $A$-type correction, such contributions are due to
pairs of paths that differ by a sequence of scattering events
traversed in opposite directions as is indicated in
Fig.~\ref{VIIfig1}.
\narrowtext
\noindent The sum of these ``maximally crossed'' segments of pairs 
of paths is represented by the building block $A$ in (\ref{VIIgdiag}).  
More specifically,
        \[
        \begin{minipage}{8.0cm}
        \psfig{figure=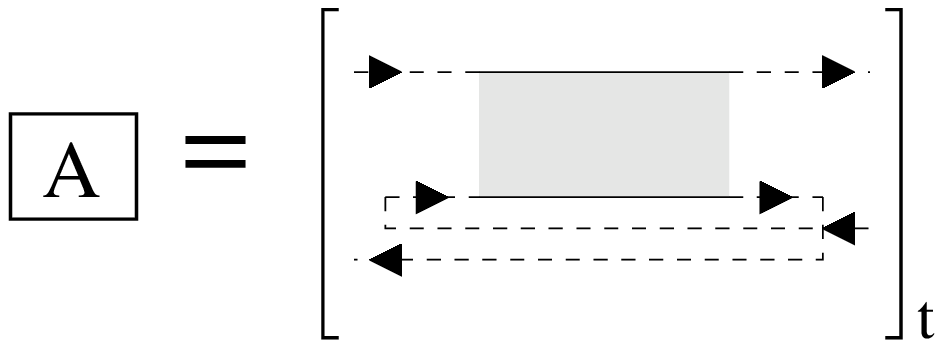,height=1.6cm}
        \end{minipage}+\dots,
        \]
where the shaded region represents an $A$-type {\it cooperon}, the
subscript `t' means that the external arrows are shown merely for 
the sake of clarity but do not contribute to the $A$-block as such, 
and the dots stand for diagrams of a more complex structure that 
have to be taken into account to obtain a result consistent with
unitarity.  It is the presence of these unitarity-preserving 
contributions that renders the calculation of the $A$-block within 
plain diagrammatic perturbation theory lengthy.  The alternative 
computational scheme presented in Appendix \ref{app1} yields
        \[
        \delta g^A = \frac{M}{2} T^2 \left( \frac{1}{MT+\eta_t}
        -\frac{3}{MT+\eta_s+\eta_t}\right),
        \]
where the parameters $\eta_s = 4N\epsilon_s$ and $\eta_t = 
4N\epsilon_t$ are the scaled symmetry-breaking parameters of our 
model.

\begin{figure}
        \centerline{\psfig{figure=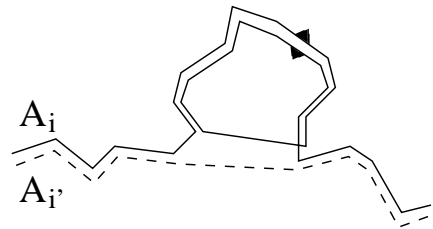,height=3.0cm}}
        \vspace{0.5cm}
        \caption{Pair of semiclassical paths contributing to the 
        $D$-type weak localization process.  The triangles represent 
        Andreev reflections.}
        \label{VIIfig2}
        \end{figure}

{\it $D$-type corrections:} A pair of paths contributing to the
$D$-type weak localization process is shown in Fig.~\ref{VIIfig2}. 
\noindent Note that the self-intersecting loop must contain a
nonvanishing even number of Andreev reflections (the figure displays
the simplest possible case of just two Andreev events).  We note in
passing that the $A$-type loop shown in Fig.~\ref{VIIfig1} may contain
Andreev reflections, too (for this reason we said that the NS $A$-type
correction is analogous to, though not identical with, the normal weak
localization correction), their presence is just not imperative like
in the $D$-case.  Clearly, the $D$-type correction does not have any
analog in normal metals.  Note also that the closed loop in
Fig.~\ref{VIIfig2} involves only one of the two paths.  This shows
that the existence of the $D$-type process is essentially due to the
nontrivial behavior of the single-particle Green's function.  The same
mechanism of quantum coherence at the single-particle level was
responsible for the correction to the single-particle density of
states discussed in Sec.~\ref{OnePoint}.

In diagrammatic language, the loop insertion in Fig.~\ref{VIIfig2} 
is represented by
        \[
        \begin{minipage}{8.0cm}
        \psfig{figure=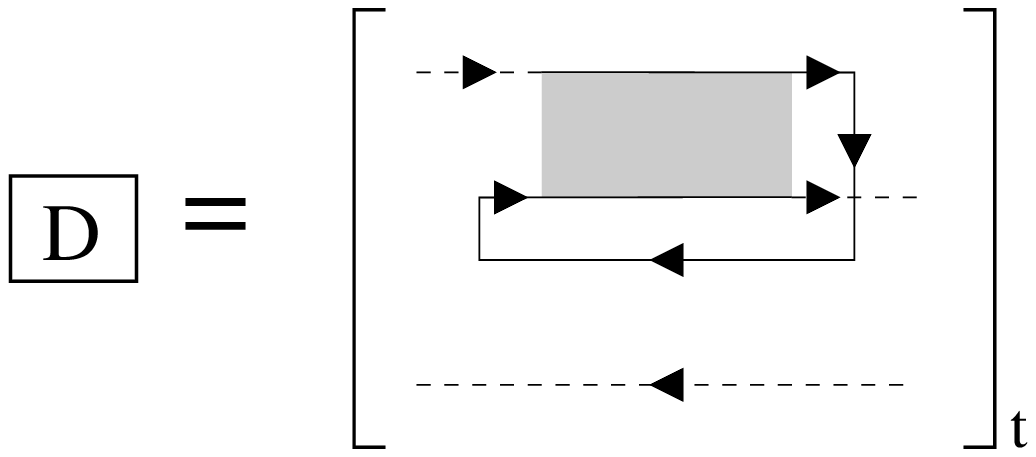,height=2.5cm}
        \end{minipage}+\dots,
        \]
where the shaded region now represents a $D$-type cooperon mode and
the dots stand for a set of unitarity-preserving counter diagrams.
The quantitative analysis yields
        \[
        \delta g^D = 
        (1-T)\left(1-\frac{3MT}{MT+\eta_s}\right).
        \]
A striking feature of this expression is its insensitivity 
to the breaking of time-reversal symmetry: the $D$-type weak 
localization correction for NS-systems survives the application 
of external magnetic fields\cite{brouwer_wl}.  Collecting terms we 
obtain the final result
        \begin{eqnarray}
        &&\langle g \rangle = 2MT
        +(1-T)\left(1-\frac{3MT}{MT+\eta_s}\right)
        \nonumber \\
        &&\hspace{0.5cm} + \frac{M}{2}T^2
        \left( \frac{1}{MT+\eta_t} - 
        \frac{3}{MT+\eta_s+\eta_t} \right)
        \nonumber \\ 
        &&\hspace{4cm} + {\cal O}(1/M,M/N)
        \label{wl_result}
        \end{eqnarray}
for the dimensionless mean conductance of our system at zero bias.  
We see that the $D$-type correction is zero for $T = 1$, while the
$A$-type correction vanishes in the limit $T \to 0$, with $MT$ held 
fixed.  From what has been said about the $D$-type modes, we expect
the $D$-type correction to disappear at finite bias.  Detailed
analysis shows that the crossover scale for this to happen is 
determined by $\nu\omega \sim MT$. 

\section{Universal Conductance Fluctuations}
\label{sec:ucf}

The conductance fluctuations of normal-conducting systems\cite{lsf}
have been studied extensively.  They are independent of system size
and strength of the disorder and depend only on symmetry.  The latter
dependence can be summarized by saying that ${\rm var}(g)$ is
proportional to the number of massless modes for a given universality
class.  When all symmetries are broken, the $A$-type spin-singlet
diffuson is the only mode which is massless.  As we switch off the
spin-orbit interaction, the $A$-type spin-triplet diffuson modes
become massless, too, which increases ${\rm var}(g)$ by a factor of
four.  If in addition time reversal is a good symmetry, the cooperon
modes become massless, thereby increasing ${\rm var}(g)$ by yet
another factor of two. 

The NS-systems considered in the present paper are
expected\cite{brouwer_ucf} to show conductance fluctuations that are
qualitatively similar to those of N-systems.  To calculate the
variance, we may use an extension of the diagrammatic method described
in the previous section or, alternatively, we may map our
random-matrix model on a zero-dimensional field theory of the
nonlinear $\sigma$ model type.  In the present paper neither of these
methods will be used.  Instead, we will turn to another approach,
which is restricted to the strong-coupling limit $T = 1$ but has the
great advantage of being very simple. 

The symmetry properties of the $S$-matrix derive from the symmetries
of the Hamiltonian by exponentiation.  As before, let $M$ denote the
number of channels in the lead, not counting spin and particle-hole
degeneracy.  By the considerations of Sec.~\ref{sec:class} the
$S$-matrix may be regarded as an element of the symmetric space ${\rm
SO}(4M)$ for class $D$, ${\rm Sp}(2M)$ for $C$, ${\rm SO}(4M)/{\rm
U}(2M)$ for $D$III, and ${\rm Sp}(2M)/{\rm U}(M)$ for $C$I.  We will
refer to these spaces as ``$S$-matrix manifolds'' for short.  Let $A =
(a,s,\sigma)$ $(a=1,...,M; s=\pm 1/2; \sigma={\rm p,h})$ be a
composite index.  From the definition of the transmission coefficient
$T$ as a ``sticking probability'' \cite{haw} we have $T = 1 - |\langle
S_{AA} \rangle|^2 + {\cal O} (1/MT)$.  Therefore $T = 1$ implies an
$S$-matrix with vanishing ensemble average, which means that $S$ can
be taken to be uniformly distributed on its $S$-matrix manifold. 

\subsection{Class $C$}

For the symmetry classes $C$ and $C$I the $S$-matrix operates on the
tensor product of channel space and particle-hole space, while spin is
accounted for by multiplication of the conductance by a factor of two.
Recall the definition of the symplectic group ${\rm Sp}(2M)$ by
      \begin{equation}
      {U^{-1}}^\dagger = U = {\cal C} {U^{-1}}^{\rm T} {\cal C}^{-1}
      \label{sp2m}
      \end{equation}
where ${\cal C} = {\bf 1}_M \otimes i \sigma_y$.  In keeping with the 
above, we take the $S$-matrix $S \equiv U$ for class $C$ to be uniformly
distributed on ${\rm Sp}(2M)$. In other words, ensemble averages 
$\langle ... \rangle$ are computed by integrating with respect to the 
Haar measure $dU$:
      \[
      \langle f(U) \rangle = \int_{{\rm Sp}(2M)} f(U) dU .
      \]
The canonical projection of ${\rm Sp}(2M)$ onto the coset space 
${\rm  Sp}(2M)/{\rm U}(M)$ by $U \mapsto U U^{\rm T}$ turns the Haar
measure of the former into the invariant (or uniform) measure of the 
latter.  Therefore, ensemble averages for class $C$I can be obtained 
from
      \[
      \langle f(S) \rangle_{C{\rm I}} = 
      \langle f(UU^{\rm T}) \rangle .
      \]
Because the Haar integral is invariant under left and right 
translations,
      \[
      \int_{{\rm Sp}(2M)} f(U_L U U_R) dU = 
      \int_{{\rm Sp}(2M)} f(U) dU ,
      \]
the defining equations for ${\rm Sp}(2M)$ lead to
      \begin{eqnarray}
      \langle U_{AB} \rangle &=& 0 ,
      \nonumber \\
      \langle U_{AB}^{\vphantom{*}} 
      U_{CD}^* \rangle
      &=& \delta_{AC} \delta_{BD} / 2M ,
      \label{inv_two} \\
      \langle U_{AB} U_{CD} \rangle
      &=& {\cal C}_{AC} {\cal C}_{BD} / 2M .
      \nonumber
      \end{eqnarray}
To compute the ensemble average of a product of two $U$'s and two
${U^*}$'s we note that if $\psi_A \in V$ are the components of a
vector transforming according to the fundamental representation of
${\rm Sp}(2M)$, there exist only two independent invariants on $V
\otimes V^* \otimes V \otimes V^*$, namely $\sum \psi_A^{\vphantom{*}}
\psi_A^* \psi_B^{\vphantom{*}}\psi_B^*$ and $\sum{\cal C}_{AB}
{\cal C}_{CD} \psi_A \psi_B \psi_C^* \psi_D^*$.  Using this elementary
group-theoretical fact we obtain
      \begin{eqnarray}
      &&\langle U_{A_1 B_1}^{\vphantom{*}} U_{C_1 D_1}^*
      U_{A_2 B_2}^{\vphantom{*}} U_{C_2 D_2}^* \rangle
      \nonumber \\
      = &&{2M-1 \over 2M(2M+1)(2M-2)} \Bigl[
      \delta_{A_1 C_1} \delta_{A_2 C_2} 
      \delta_{B_1 D_1} \delta_{B_2 D_2}
      \nonumber \\
      &&\qquad\qquad\qquad\qquad\qquad
      +\delta_{A_1 C_2} \delta_{A_2 C_1} 
      \delta_{B_1 D_2} \delta_{B_2 D_1}
      \nonumber \\
      &&\qquad\qquad\qquad\qquad\qquad
      +{\cal C}_{A_1 A_2} {\cal C}_{C_1 C_2}
      {\cal C}_{B_1 B_2} {\cal C}_{D_1 D_2} \Bigr]
      \nonumber \\
      - &&{1 \over 2M(2M+1)(2M-2)} \Bigl[
      \delta_{A_1 C_1} \delta_{A_2 C_2} 
      \delta_{B_1 D_2} \delta_{B_2 D_1}
      \label{inv_four} \\
      &&\qquad\qquad\qquad\qquad\qquad
      +\delta_{A_1 C_2} \delta_{A_2 C_1} 
      \delta_{B_1 D_1} \delta_{B_2 D_2}
      \nonumber \\
      &&\qquad\qquad
      +(\delta_{A_1 C_1} \delta_{A_2 C_2} -
      \delta_{A_1 C_2} \delta_{A_2 C_1} )
      {\cal C}_{B_1 B_2} {\cal C}_{D_1 D_2}
      \nonumber \\
      &&\qquad\qquad 
      +{\cal C}_{A_1 A_2} {\cal C}_{C_1 C_2} 
      (\delta_{B_1 D_1} \delta_{B_2 D_2} - 
      \delta_{B_1 D_2} \delta_{B_2 D_1}) \Bigr] .
      \nonumber
      \end{eqnarray}
The numerical coefficients in this expression are determined by
summing over any two pairs of equal indices and then comparing the
results to (\ref{inv_two}) using the relations (\ref{sp2m}).  
Equation (\ref{inv_four}) entails
      \[
      \sum_{AB} \langle U_{{\rm p}A}^{\vphantom{*}} 
      U_{{\rm h}A}^{\vphantom{*}} U_{{\rm p}B}^* U_{{\rm h}B}^* 
      \rangle = (2M+1)^{-1} ,
      \]
which can be used to compute the weak localization correction for 
class $C$I.  Summing over initial and final (or particle and hole)
channels, we get $\langle{\rm Tr}S^{\rm ph} S^{\dagger{\rm hp}} 
\rangle_{C{\rm I}} = M^2 / (2M+1)$ which yields $\delta g = - 1$ 
in agreement with (\ref{wl_result}).  To calculate the conductance 
fluctuations for class $C$, we deduce from (\ref{inv_four})
      \[
      \langle \left( {\rm Tr} S^{\rm ph} S^{\dagger{\rm hp}}
      \right)^2 \rangle = {M^2 \over 4} + {1\over 8} 
      + {\cal O}(M^{-1}) .
      \]
Subtracting the square of the first moment and multiplying by a 
factor of $4\times 4$ for charge and spin, we get ${\rm var}(g) = 2$. 

\subsection{Class $D$}

The symmetry class $D$ can be treated by direct transcription from
class $C$, the only difference being the way the spin enters.  The
$S$-matrix now operates on the full tensor product of channel space,
particle-hole space and spin space.  The $S$-matrix manifold for $D$
is isomorphic to the orthogonal group ${\rm SO}(4M)$, and is defined
by (\ref{sp2m}) with ${\cal C} = {\bf 1}_M \otimes \sigma_x \otimes
{\bf 1}$.  Equations (\ref{inv_two}) remain formally unchanged except
for the replacement $2M \to 4M$.  The projection $U \mapsto U\tau
U^{\rm T}\tau^{-1}$ with $\tau = {\bf 1}_M \otimes {\bf 1} \otimes
i\sigma_y$ takes the Haar measure of ${\rm SO}(4M)$ into the invariant
measure of ${\rm SO}(4M)/{\rm U}(2M)$.  Ensemble averages are given by
      \begin{eqnarray}
      \langle f(S) \rangle_{D{\rm III}}
      &=& \langle f(U\tau U^{\rm T}\tau^{-1}) \rangle ,
      \nonumber \\
      \langle f(U) \rangle 
      &=& \int_{{\rm SO}(4M)} f(U) dU.
      \nonumber
      \end{eqnarray}
The ensemble average of a product of four $U's$ is
      \begin{eqnarray}
      &&\langle U_{A_1 B_1}^{\vphantom{*}} U_{C_1 D_1}^*
      U_{A_2 B_2}^{\vphantom{*}} U_{C_2 D_2}^* \rangle
      \nonumber \\
      = &&{4M+1 \over 4M(4M-1)(4M+2)} \Bigl[
      \delta_{A_1 C_1} \delta_{A_2 C_2} 
      \delta_{B_1 D_1} \delta_{B_2 D_2}
      \nonumber \\
      &&\qquad\qquad\qquad\qquad\qquad
      +\delta_{A_1 C_2} \delta_{A_2 C_1} 
      \delta_{B_1 D_2} \delta_{B_2 D_1}
      \nonumber \\
      &&\qquad\qquad\qquad\qquad\qquad
      +{\cal C}_{A_1 A_2} {\cal C}_{C_1 C_2}
      {\cal C}_{B_1 B_2} {\cal C}_{D_1 D_2} \Bigr]
      \nonumber \\
      - &&{1 \over 4M(4M-1)(4M+2)} \Bigl[
      \delta_{A_1 C_1} \delta_{A_2 C_2} 
      \delta_{B_1 D_2} \delta_{B_2 D_1}
      \nonumber \\
      &&\qquad\qquad\qquad\qquad\qquad
      +\delta_{A_1 C_2} \delta_{A_2 C_1} 
      \delta_{B_1 D_1} \delta_{B_2 D_2}
      \nonumber \\
      &&\qquad\qquad
      +(\delta_{A_1 C_1} \delta_{A_2 C_2} +
      \delta_{A_1 C_2} \delta_{A_2 C_1} )
      {\cal C}_{B_1 B_2} {\cal C}_{D_1 D_2}
      \nonumber \\
      &&\qquad\qquad 
      +{\cal C}_{A_1 A_2} {\cal C}_{C_1 C_2} 
      (\delta_{B_1 D_1} \delta_{B_2 D_2} +
      \delta_{B_1 D_2} \delta_{B_2 D_1}) \Bigr] .
      \nonumber
      \end{eqnarray}
The remaining calculations are the same as before.  We obtain
$\delta g = + 1/2$ for class $D$III, and ${\rm var}(g) = 1/2$ 
for class $D$. 

\subsection{Conjecture for $C$I and $D$III}

For N-systems the breaking of time-reversal symmetry is known to
reduce ${\rm var}(g)$ by a factor of two, while the breaking of
spin-rotation invariance causes a reduction by a factor of four.  As
was said earlier, this pattern is explained by the observation that
${\rm var}(g)$ simply counts the number of massless modes in each
universality class.  From our experience with diagrammatic
perturbation theory of the model (\ref{correlator},\ref{VIISmat}) we
expect the same principle to be operative here, i.e. we expect ${\rm
var}(g)$ to be still determined by the number of massless modes.
Indeed, the conductance fluctuations for the classes $C$ and $D$ are
seen to be bigger than the corresponding fluctuations for N-systems by
a factor of eight.  To understand this, we note that there is a
trivial enhancement by a factor of $2^2 = 4$ due to the transfer of
{\it two} elementary charges in an Andreev reflection.  The other
factor of two can be interpreted as telling us that the number of
massless modes a priori is twice as large: for every $A$-type mode,
which is already present in the N-system, there exists an extra
$D$-type (or BdG particle-hole) mode in the NS-system, see
Sec.~\ref{sec:modes}.  By extrapolation we are led to the following
{\it conjecture}:
      \[
      {\rm var}(g) = \left\{ \matrix{
      4 &\quad (C{\rm I}), \cr
      2 &\quad (C), \cr
      1 &\quad (D{\rm III}), \cr
      1/2 &\quad (D) , \cr} \right.
      \]
which differs from the result of Brouwer and Beenakker\cite{brouwer_ucf}
who found the size of the conductance fluctuations to depend only
weakly on whether time-reversal symmetry is broken or not.  Note
however that their result applies to a different situation than the
one considered here (In their case the superconducting order parameter
is homogeneous in space for class $C$I).

\section{Conclusions}
\label{sec:conc}

In this paper we have initiated the study of a special family of
NS-systems where the spatial variation of the superconducting order
parameter is such that the Andreev phase shift averages to zero along
a typical semiclassical single-electron trajectory.  We find such
systems particularly interesting because the proximity effect is 
inoperative and quasiparticle states exist right at the chemical
potential.  Disorder or dynamically generated chaos mixes the states
and leads to a novel and universal type of level statistics within an
energy window whose size is determined by the frequency of Andreev
reflection.  By classifying systems according to their symmetries we
identified four universality classes, denoted by $C$, $C$I, $D$, and
$D$III.  Time reversal is a good (broken) symmetry for $C$I and $D$III
($C$ and $D$), while spin is conserved (not conserved) for $C$ and
$C$I ($D$ and $D$III).  For each universality class the joint
probability distribution of the quasiparticle energy levels was given
in closed form.  The $n$-level correlation functions for the classes
$C$ and $D$ were calculated by the mapping onto a free Fermi gas on
a half-line with  Dirichlet and Neumann boundary conditions at the 
origin.  The joint probability distributions of the levels for $C$I 
and $D$III were transformed into those of the Laguerre Orthogonal and 
Laguerre Symplectic Ensembles, whose level statistics has been worked 
out completely (albeit with a minor computational error) by Nagao and
Slevin. 

To calculate the transport properties of open systems in the
zero-dimensional limit, we formulated a random-matrix model and
treated it using a variant of the impurity diagram technique.  An
important feature we pointed out was the doubling of the number of
low-energy modes in comparison with conventional normal-conducting
systems.  For every $A$-type mode, i.e. for every BdG particle-particle
(or hole-hole) spin-singlet or spin-triplet diffuson or cooperon,
there exists precisely one corresponding BdG particle-hole or $D$-type
mode.  The weak localization correction to the average conductance for
an NSS-geometry was calculated as a function of the ``sticking
probability'' $T$ and two perturbations breaking time-reversal
symmetry and spin-rotation invariance.  The technically more involved
task of calculating the variance of the fluctuating conductance was
carried out only for $T = 1$ and the universality classes $C$ and $D$,
by using an $S$-matrix formalism \'a la Mello.  We found ${\rm
var}(g)$ to be enhanced by a factor of two relative to the rule ${\rm
var}(g_{\rm NS}) = 4 {\rm var}(g_{\rm N})$.  We attribute this
enhancement to the doubling of low-energy modes by the coupling to the
superconductor.  Let us emphasize that the effects we have studied are
universal (in the ergodic limit) and are independent of such
microscopic detail as the NS-barrier transmittency. 

Clearly, the present paper constitutes only a first step into a new
and exciting research area of mesoscopic physics, and much more is yet
to be done.  Some of the open problems are the following.  (i) We have
shown how to solve the level statistics problem for each universality
class but more generally one might also be interested in the crossover
between classes.  Here the crossovers $C{\rm I} \to C$ and $D{\rm III}
\to D$ look amenable to analytical techniques, since the level
statistics for $C$ and $D$ (just as for the Gaussian Unitary Ensemble)
maps on a free Fermi gas problem.  (ii) Our results for the level
statistics are restricted to an energy range proportional to the
inverse mean time spent between successive Andreev reflections.  To
access the short-time or high-energy regime beyond the crossover
scale, our maximum-entropy ensembles need to be modified by allowing
for different variances of the random pairing and normal matrix
elements.  (iii) Although we have outlined the semiclassical
interpretation of the $D$-type (or particle-hole) modes, a more
detailed discussion of their role in semiclassical periodic-orbit
theory would certainly be desirable.  (iv) We need to extend our
results for the universal conductance fluctuations to the classes $C$I
and $D$III and to arbitrary $T$.  (v) While the zero-dimensional (or
ergodic) limit is adequately described by the maximum-entropy ansatz,
the diffusive and ballistic regimes necessitate a more detailed
modelling.  In particular, the nonrandom nature of the magnitude of
the pairing field will make itself felt at short times.  It is
an open technical problem how to deal analytically with the phase
randomness of Hamiltonian matrix elements when their magnitude is to
be kept fixed.  (vi) We have concentrated on an NS-geometry that is
particularly easy to treat but future work will have to include other
geometries.  (vii) Last but not least, we need to address the
nontrivial question: how large is the effect of residual Coulomb
interactions on the $D$-type modes? There is no doubt that the
short-time physics can be adequately described in a simple
independent-quasiparticle picture, but at large times the coherence
between particles and holes will get cut off by Coulomb blockade and
other correlation effects.  The question is what is the time scale
where this happens.

We shall end on a mathematical note.  According to Cartan, there exist
eleven large families of symmetric spaces.  Those of type II are the
compact unitary, orthogonal and symplectic Lie groups $(A, B, C, D)$.
The large families of type-I symmetric spaces are denoted by $A$I,
$A$II, $A$III, $BD$I, $C$I, $C$II, and $D$III.  The standard
Wigner-Dyson universality classes derive from $A$ (GUE), $A$I (GOE),
and $A$II (GSE), while the universality classes of a massless Dirac
particle derive from $A$III (chGUE), $BD$I (chGOE), and $C$II (chGSE).
As we have shown, the new universality classes found in mesoscopic
NS-systems exhaust the remaining large families of symmetric spaces
except for $B$ (the orthogonal group in odd dimensions), which does
not occur.  Thus, if we group $B$ together with $D$, there is a {\it
bijection} between the known universality classes of disordered
single-particle systems and the large families of symmetric spaces.
We consider this to be a strong indication that no other universality
classes will be found, since any additional universality class would
exceed Cartan's scheme and therefore would have to be of a different
nature. 

{\bf Acknowledgment.} This work was supported in part by the Deutsche
Forschungsgemeinschaft, SFB 341.

\appendix
\section{Conductance and Ward Identities}
\label{app1}

In this
appendix we elaborate on the calculation of the weak localization
correction to the conductance, Eq.~(\ref{wl_result}).  Owing to the
presence of the BdG particle-hole degree of freedom, this calculation
turns out to be much more involved than in pure N-systems.  For this
reason we prefer to control the diagrammatic expansion by means of
exact algebraic relationships.  The basic concepts used in this
appendix have been introduced in the seminal paper\cite{vw}.

To begin with, we recapitulate two Ward identities that will play a
crucial role in what follows. Let us write the average retarded
Green's function as $\langle G_{\alpha,\alpha'}\rangle =
\delta_{\alpha, \alpha'} (-\Sigma_{\mu}+i\Gamma_{\mu})^{-1} =:
\delta_{\alpha, \alpha'} G_{\mu}$, where $\alpha=(\mu,s,q)$ is a
composite index accounting for the site ($\mu)$, spin ($s$) and
particle-hole ($q = {\rm p,h}$) degrees of freedom.  (After averaging,
the Green's function depends only on the site index.)  The first Ward
identity is immediate from the definitions and reads
        \begin{equation}
        G_{\mu}^{\vphantom{*}} G_{\mu}^*=-\Delta G_{\mu}
        \left(-\Delta\Sigma_{\mu}+2i\Gamma_{\mu}\right)^{-1},
        \label{Awardone}
        \end{equation}
where $G_{\mu}^*$ is the average advanced Green's function,
$\Delta G_{\mu}=G_{\mu}-G_{\mu}^*$ and $\Delta\Sigma_{\mu} =
\Sigma_{\mu} - \Sigma_{\mu}^*$.  A second and less trivial
Ward-identity\cite{vw} relates the self energy $\Sigma$ to the
so-called irreducible two-particle vertex $U$:
        \begin{equation}
        \Delta\Sigma_{\mu}=\sum_{\alpha'}
        U_{\alpha,\alpha'}\Delta G_{\mu'}.
        \label{Awardtwo}
        \end{equation}
The irreducible vertex $U$ is defined as the set of all truncated 
four-point functions that cannot be cut by just cutting two average 
Green's functions, i.e.
        \[
        U=
        \begin{minipage}{5.0cm}
        \psfig{figure=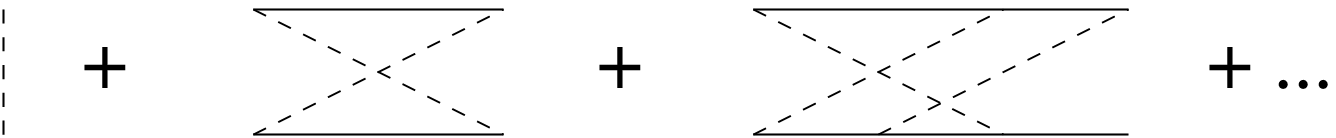,height=0.7cm}
        \end{minipage}.
        \]
In the following we focus on the analysis of the auxiliary quantity
        \[
        \Phi_{\alpha}= \sum_{\alpha' ; q'={\rm p}} \langle 
        \left|G_{\alpha,\alpha'}\right|^2 \rangle \Gamma_{\mu'}.
        \]
>From this the mean conductance is obtained as (cf. (\ref{VIIggreen})) 
        \begin{equation}
        \langle g \rangle 
        = 8 \sum_{\alpha;q={\rm h}} \Gamma_\mu \Phi_{\alpha}.
        \label{Acond}
        \end{equation} 
We start from the ansatz
        \begin{eqnarray}
        &&\Phi_{\alpha}=\Delta
        G_{\mu}\Big[c+c_s(-)^s+c_q(-)^q+c_{sq}(-)^{s+q}+
        \nonumber\\
        &&\hspace{1.0cm}+
        \Gamma_{\mu}\left(d+d_s(-)^s+d_q(-)^q
        +d_{sq}(-)^{s+q}\right)\Big].
        \end{eqnarray}
(Although other expressions involving arbitrary functions of
$\Gamma_{\mu}$ seem possible, this formula does represent the most 
general starting point.  Equation (\ref{VIIWdiagonal}) yields
$\sum_{\mu}\Gamma_{\mu}^n = \gamma^{n-1}\sum_{\mu} \Gamma_{\mu}$,
so it is sufficient to start from an expression that is linear in
the coefficients $\Gamma_{\mu}$.) The quantity $\Phi$ satisfies
Dyson's equation\cite{vw}
        \begin{equation}
        \left(\Delta\Sigma_{\mu}-2i\Gamma_{\mu}\right)\Phi_{\alpha}
        = \Delta G_{\mu}\left(\Gamma_{\mu}\delta_{q{\rm p}}+
        \sum_{\alpha'} U_{\alpha,\alpha'}\Phi_{\alpha'}\right)
        \hspace{0.3cm}(*),
        \label{ADyson}
        \end{equation}
where use of the identity (\ref{Awardone}) has been made. To fix the
coefficients $c,\dots,d,\dots$ we subject Eq.~(\ref{ADyson}) to
various summation procedures.  For example, by taking the sum
$\sum_{\alpha}(*)$ and then using the second Ward identity
(\ref{Awardtwo}), we obtain
        \begin{equation}
        c + \gamma d = -1/4i .
        \label{Ares1}
        \end{equation}
Seven more equations for the remaining coefficients are generated by
performing the summations $\sum_{\alpha}(-)^s(*)$,
$\sum_{\alpha}(-)^q(*)$, $\sum_{\alpha}(-)^{s+q}(*)$,
$\sum_{\alpha}\Gamma_{\mu}(*)$, $\sum_{\alpha}\Gamma_{\mu}(-)^s(*)$,
$\sum_{\alpha}\Gamma_{\mu}(-)^q(*)$ and
$\sum_{\alpha}\Gamma_{\mu}(-)^{s+q}(*)$. The outcome of all this
may be cast in the form of a matrix equation
        \begin{equation}
        \left(\begin{array}{cc}
        A&B\\
        C&D
        \end{array}\right)
        \left(\begin{array}{c}
        \vec{c}\\
        \vec{d}
        \end{array}\right)=-2a_1
        \left(\begin{array}{c}
        \vec{u}\\
        \vec{v}
        \end{array}\right),
        \label{Amatrix}
        \end{equation}
where $\vec{c}^{\rm T} = (c_s,c_q,c_{sq})$, $\vec{d}^{\rm T} =
(d,d_s,d_q,d_{sq})$, $\vec{u}^{\rm T}=(0,1,0)$, $\vec{v}^{\rm
T}=(0,0,\gamma,0)$, and $a_1 = \sum_{\mu}\Gamma_{\mu}\Delta
G_{\mu}$. Fortunately, it is easy to invert the $7\times7$
matrix appearing in this equation.  By construction, the
coefficients appearing in the subblock $A(B,C,D)$ involve summations
over the index $\alpha$ that do not (do) contain matrix elements
$\Gamma_{\mu}$.  Since $\sum_{\alpha}\dots \sim{\cal O}(N)$, whereas
$\sum_{\alpha}\Gamma_{\mu}\dots \sim{\cal O}(M)$, the coefficients
appearing in $A$ exceed those in the remaining subblocks by a large
factor of ${\cal O}(N/M)$. We thus conclude
        \begin{equation}
        \left(\begin{array}{cc}
        A&B\\
        C&D
        \end{array}\right)^{-1}\simeq
        \left(\begin{array}{cc}
        0&0\\
        0&D^{-1}
        \end{array}\right)\Rightarrow
        \vec{c}\simeq0, \ \vec{d}\simeq D^{-1}\vec{v}.
        \label{Ares2}
        \end{equation}
The matrix $D$ can easily be inverted as it is already of diagonal
form.  Combining Eqs.~(\ref{Acond}), (\ref{Ares1}) and (\ref{Ares2}),
we obtain
        \begin{equation}
        \langle g \rangle =16a_1\left(-\frac{1}{4i}-
        \frac{2\gamma^2 a_1}{4\gamma c_1-8i\gamma^2 
        a_1-\overline{U}}\right),
        \end{equation}
where $c_1=\sum_{\mu}\Gamma_{\mu}\Delta G_{\mu}\Delta\Sigma_{\mu}$ and
        \[
        \overline{U}=\sum_{\alpha,\alpha'}
        \Gamma_{\mu}\Delta G_{\mu}(-)^q
        U_{\alpha,\alpha'}
        \Gamma_{\mu'}\Delta G_{\mu'}(-)^{q'}.
        \]
We next decompose the self energy according to $\Sigma_{\mu}=
-i/\lambda+\delta\Sigma_{\mu}$ into a leading order contribution plus
a correction term $\delta\Sigma_{\mu}$ of ${\cal O}(1/M)$.
Anticipating that the terms $\delta\Sigma_{\mu}$ and $\overline{U}$
are of the same order, we obtain the preliminary result
        \begin{eqnarray}
        &&\langle g \rangle =
        2MT+2MT\frac{\lambda-\gamma}{\lambda(\lambda+\gamma)}
        {\rm Im}\overline{\delta\Sigma}+\frac{1}{2(\lambda+\gamma)^2}
        \overline{U}+\nonumber\\
        &&\hspace{0.5cm}+{\cal O}(1/M),
        \label{Agfinal}
        \end{eqnarray}
where $\overline{\delta\Sigma}=(M\gamma)^{-1}\sum_{\mu}\Gamma_{\mu}
\delta\Sigma_{\mu}$.  We next analyze the building blocks
$\overline{\delta\Sigma}$ and $\overline{U}$ by diagrammatic
methods.  Because Eq.~(\ref{Agfinal}) is based on Ward identities, it
automatically incorporates the condition of unitarity.  As a
consequence, the following calculation is much simpler than a direct
diagrammatic analysis of Eq.~(\ref{VIIggreen}).

To leading order in $M^{-1}$, the self-energy correction 
$\overline{\delta\Sigma}$ is given by
        \begin{equation}
        \overline{\delta\Sigma} = 
        \frac{1}{4M\gamma}\sum_{\alpha}\Gamma_{\mu}
        \begin{minipage}{7.0cm} 
        \psfig{figure=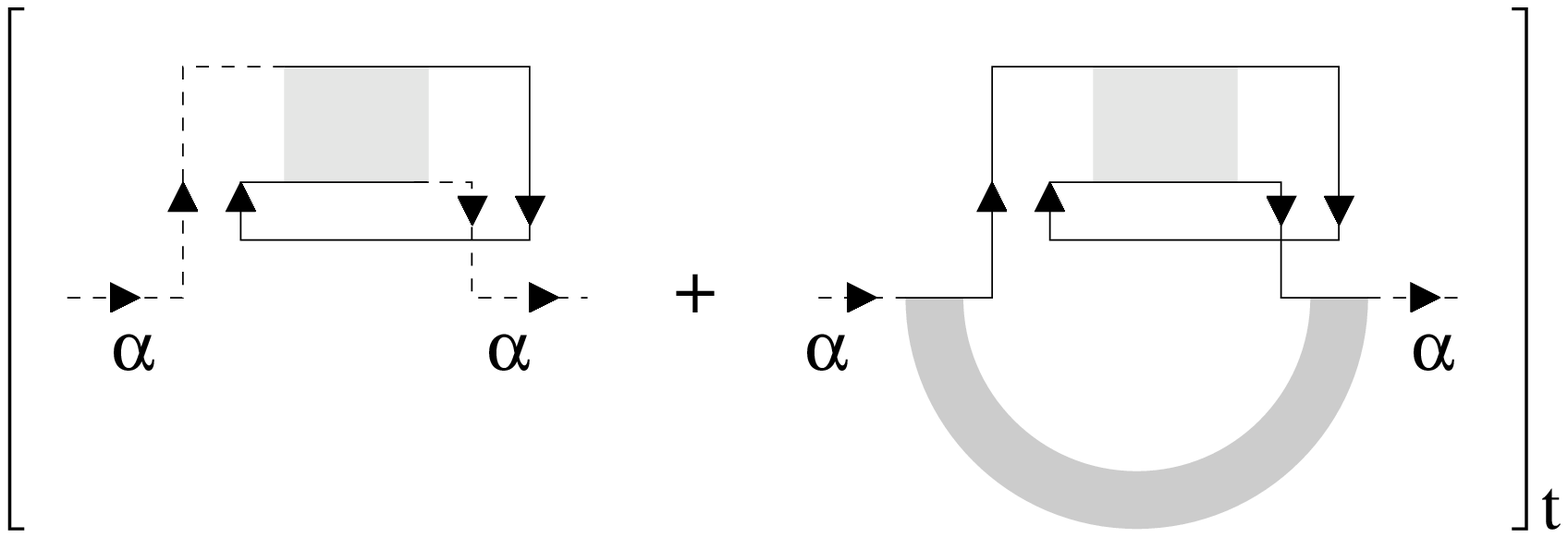,height=1.8cm}
        \end{minipage},
        \end{equation}
where the light-(dark-)shaded region represents a $D$--type
cooperon mode (a non--singular $\Pi_A^{d0}$--ladder). 
Evaluation of the diagrams yields (cf. the explanation in connection
with Fig.\ref{fig2})
        \begin{equation}
        {\rm Im} \overline{\delta\Sigma} = {\lambda(\lambda-\gamma)
        \over 2(\lambda+\gamma)}
        \left[\frac{1}{MT}-\frac{3}{MT+\eta_s}\right].
        \label{Adelsig}
        \end{equation}
The dominant contribution to the vertex correction $\overline{U}$
results from an $A$--type cooperon:
        \begin{eqnarray}
        &&\overline{U}=\sum_{\alpha,\alpha'}(-)^q\Gamma_{\mu}\Delta G_{\mu}
        \begin{minipage}{2.5cm}
        \psfig{figure=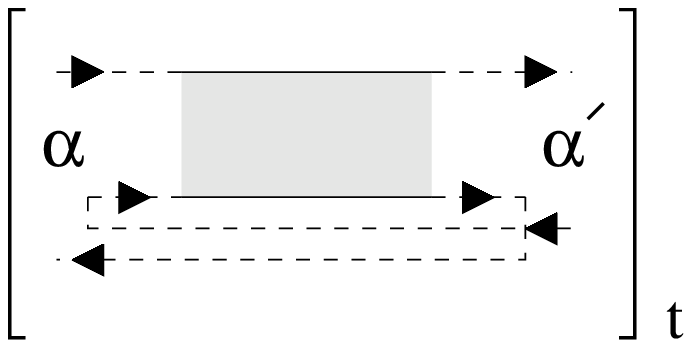,height=1.4cm}
        \end{minipage}
        \Delta G_{\mu'}\Gamma_{\mu'}(-)^{q'}=
        \nonumber\\
        &&\hspace{0.5cm}=4\lambda\gamma TM\left(\frac{1}{MT+\eta_t}
        -\frac{3}{MT+\eta_s+\eta_t}\right).\nonumber\\
        \label{Ad}
        \end{eqnarray}
By combining Eqs.~(\ref{Agfinal}), (\ref{Adelsig}) and (\ref{Ad}), and
using the expression for the transmission coefficient $T = 4 \lambda
\gamma / (\lambda+\gamma)^2$, we arrive at the final result given in
Sec.~\ref{sec:wl}.

\end{multicols}

\begin{thebibliography}{99}
\bibitem{footnote1} Note that this expansion scheme is analogous to
the $(k_F l)^{-1}$-expansion ($k_F$ is the Fermi momentum and $l$
the elastic mean free path due to impurity scttering) commonly
adopted in mesoscopic physics.  For examples, in a sample of 
extension $\sim l$ coupled to leads of comparable width, 
$N\sim(k_F l)^d$ and $M\sim (k_F l)^{d-1}$.
\bibitem{footnote2} To be general, we should consider a set of 
channel-dependent transmission coefficients $\{T_a\}$.  However, since 
this generalization does not lead to qualitatively different results 
but only complicates the notation, we put $T_a = T$ independent of $a$.
\end{thebibliography}
\end{document}